%% file: Blanco1_EB.tex
\documentclass[fleqn,usenatbib]{mnras}

\usepackage[T1]{fontenc}

\usepackage{graphicx}	
\usepackage{amsmath}	
\usepackage{amssymb}	
\usepackage{newtxtext,newtxmath}

\input{commands.tex}
\input{properties.tex}

\title[NCS III: A low-mass EB in Blanco 1]{NGTS clusters survey -- III: A low-mass eclipsing binary in the Blanco 1 open cluster spanning the fully convective boundary}

\author[G. Smith et al.]{
\parbox{\textwidth}{
Gareth D. Smith,$^{1}$\thanks{E-mail: \href{gds38@cam.ac.uk}{gds38@cam.ac.uk}}
Edward Gillen,$^{2,1}$\thanks{Winton Fellow}
Didier Queloz,$^{1}$
Lynne A. Hillenbrand,$^{3}$
Jack S. Acton,$^{4}$
Douglas R. Alves,$^{11}$
David~R.~Anderson,$^{5,6}$
Daniel Bayliss,$^{5,6}$
Joshua~T.~Briegal,$^{1}$
Matthew~R.~Burleigh,$^{4}$
Sarah~L.~Casewell,$^{4}$
Laetitia Delrez,$^{7,8}$
Georgina Dransfield,$^{9}$
Elsa Ducrot,$^{8}$
Samuel Gill,$^{5,6}$
Micha\"el Gillon,$^{8}$
Michael~R.~Goad,$^{4}$
Maximilian N. G{\"u}nther,$^{10}$\thanks{Juan Carlos Torres Fellow} 
Beth~A.~Henderson,$^{4}$
James~S.~Jenkins,$^{11,12}$
Emmanu\"el Jehin,$^{7}$
Maximiliano~Moyano,$^{13}$
Catriona A. Murray,$^{1}$
Peter P. Pedersen,$^{1}$
Daniel Sebastian,$^{9}$
Samantha Thompson,$^{1}$
Rosanna~H.~Tilbrook,$^{4}$
Amaury~H.M.J.~Triaud,$^{9}$
Jose~I.~Vines,$^{11}$
Peter J. Wheatley$^{5,6}$
}
\\
$^{1}$Astrophysics Group, Cavendish Laboratory, J.J. Thomson Avenue, Cambridge CB3 0HE, UK\\
$^{2}$Astronomy Unit, Queen Mary University of London, Mile End Road, London E1 4NS, UK\\
$^{3}${California Institute of Technology, Department of Astronomy; 1200 East California Blvd; Pasadena, CA, 91125, USA}\\
$^{4}$School of Physics and Astronomy, University of Leicester, University Road, Leicester, LE1 7RH, UK\\
$^{5}$ Department of Physics, University of Warwick, Gibbet Hill Road, Coventry CV4 7AL, UK \\
$^{6}$ Centre for Exoplanets and Habitability, University of Warwick, Gibbet Hill Road, Coventry CV4 7AL, UK \\ 
$^{7}$Space sciences, Technologies and Astrophysics Research (STAR) Institute, University of Li\`ege, Belgium\\
$^{8}$Astrobiology Research Unit, University of Li\`ege, All\'ee du 6 ao\^ut, 19, 4000 Li\`ege (Sart-Timan), Belgium \\
$^{9}$School of Physics \& Astronomy, University of Birmingham, Edgbaston, Birmingham, B15 2TT, UK\\
$^{10}$Department of Physics, and Kavli Institute for Astrophysics and Space Research, Massachusetts Institute of Technology, Cambridge, MA 02139, USA\\
$^{11}$Departamento de Astronomia, Universidad de Chile, Casilla 36-D, Santiago, Chile\\
$^{12}$N\'ucleo de Astronom\'ia, Facultad de Ingenier\'ia y Ciencias, Universidad Diego Portales, Av. Ej\'ercito 441, Santiago, Chile\\
$^{13}$Instituto de Astronom\'ia, Universidad Cat\'{o}lica del Norte, Angamos 0610, 1270709, Antofagasta, Chile
}

\date{Accepted XXX. Received YYY; in original form ZZZ}

\pubyear{2021}

\begin{document}
\label{firstpage}
\pagerange{\pageref{firstpage}--\pageref{lastpage}}
\maketitle

\begin{abstract}
We present the discovery and characterisation of an eclipsing binary identified by the Next Generation Transit Survey in the $\sim$115 Myr old Blanco 1 open cluster. \Nsys{} comprises three M dwarfs: a short-period binary and a companion in a wider orbit. This system is the first well-characterised, low-mass eclipsing binary in Blanco 1. With a low mass ratio, a tertiary companion and binary components that straddle the fully convective boundary, it is an important benchmark system, and one of only two well-characterised, low-mass eclipsing binaries at this age. We simultaneously model light curves from NGTS, \tess, SPECULOOS and SAAO, radial velocities from VLT/UVES and Keck/HIRES, and the system's spectral energy distribution. We find that the binary components travel on circular orbits around their common centre of mass in $P_{\rm orb} =\period$ days, and have masses  $M_{\rm pri}=\Mpri$ M$_{\odot}$ and $M_{\rm sec}=\Msec$ M$_{\odot}$, radii $R_{\rm pri}=\Rpri$ R$_{\odot}$ and $R_{\rm sec}=\Rsec$ R$_{\odot}$, and effective temperatures $T_{\rm pri}=\Tpri$ K and $T_{\rm sec}=\Tsec$ K. We compare these properties to the predictions of seven stellar evolution models, which typically imply an inflated primary. The system joins a list of 19 well-characterised, low-mass, sub-Gyr, stellar-mass eclipsing binaries, which constitute some of the strongest observational tests of stellar evolution theory at low masses and young ages. 
\end{abstract}




\section{Introduction}
\label{sec:intro}
Theories of stellar evolution are integral to our understanding of observational astrophysics. By considering the relevant physics and phenomena, e.g. thermodynamics, ionisation states, nuclear reaction pathways, radiative transfer, convection, atmospheric opacity, interior-atmosphere boundary conditions, gravitational contraction, rotation and magnetic fields, such theories can be used to model the temporal evolution of stellar properties (radius, luminosity, effective temperature) for stars of given mass and metallicity. 

Stellar evolution models also play a role in calibrating various astrophysical time scales and relations, including the initial mass function \citep{Hillenbrand2008, Bastian2010}; the lifetimes of protoplanetary disks \citep{Haisch2001, Ribas2014}; the formation and migration timescales of giant planets \citep{Bell2013, Ribas2015}; and the age-activity-rotation relations of stars \citep{Marmajek2008, Meibom2015}. In addition, our knowledge of the true properties and occurrence rates of exoplanets hinges on our knowledge of their host stars, which depends on accurate stellar models \citep{Gaidos2013, Burke2015, Berger2018}. Given the reach of stellar evolution theory, it is essential to critically test model predictions against observation.

Open clusters act as important astrophysical testing grounds for theory, permitting the study of stellar evolution on coeval populations of stars. Detached, double-lined eclipsing binaries (EBs) in open clusters are excellent tools for calibrating evolutionary models, as they allow, through combination of photometry and spectroscopy, precise determination of masses, radii, luminosities and temperatures \citep{Andersen1991, Torres2010}. In the best cases, such as with space-based photometry, it is possible to measure mass and radius with minimal theoretical assumptions to a precision better than 1\% (e.g. \citealt{Torres2018, Maxted2020, Murphy2020, southworth2021rediscussion}). When accompanied by knowledge of cluster metallicity, which may be derived spectroscopically, EB measurements and stellar evolution models can provide age estimates (e.g. \citealt{David2019, Gillen2020EB}) and valuable comparisons between dating methods such as isochrone fitting, the lithium depletion boundary, gyrochronology and asteroseismology \citep{Soderblom2010, Soderblom2014}.

Discrepancies between observations and model predictions tend to be more common at young ages and low masses, where models have often been found to underpredict stellar radii for a given mass and overpredict effective temperatures \citep{Irwin2011, VonBraun2012, Torres_2013, Zhou2015, Dittmann2017, Triaud2020}. The extent to which these disagreements are due to model inaccuracies versus unaccounted-for systematic uncertainties related to starspots, the effects of additional stars in the system or other factors is, however, an active area of discussion \citep{Morales2010, Windmiller2010, Feiden2012, Somers2015}. 

One explanation for the departure of models from observation is that strong magnetic fields could inhibit the outward flow of energy by suppressing global convection \citep{Mullan2001} or inducing greater spot coverage \citep{Chabrier_2007}. In order to maintain balance between energy released in the core and flux leaving the stellar surface, the star would expand with an accompanying drop in effective temperature. The fact that most well-studied EBs have short orbital periods with synchronised rotational periods means that rapid rotation is likely to drive strong magnetic fields in these systems and hence could contribute to the observed discrepancies. In support of this idea, \citet{Kraus2011} found that EBs with orbital periods $\leq$1 day were elevated in the mass--radius plane. However, in their study of the radius discrepancy in low-mass stars, \citet{Spada2013} found that, although the components of short orbital period systems are seen to be the most deviant among EBs, those deviations are matched by the single-star sample. Interferometric angular diameter measurements by \citet{Boyajian_2012} led to a similar conclusion, indicating that, for a given mass, single and binary star radii are indistinguishable. \citet{Mann15} found, in their study of 183 nearby K7--M7 single stars, comparable discrepancies with models as found for EBs, suggesting that underlying model assumptions to do with opacity or convective mixing length are more likely to be the root cause. Finally, in their exploration of the radius inflation problem for M dwarfs on the main sequence, \citet{Morrell2019} took an all-sky sample of >15\,000 stars and employed a spectral energy distribution (SED) fitting method, using \gaia{} DR2 distances and multiwave-band photometry, to determine empirical relations between luminosity, temperature and radius. They found measured radii to be inflated by 3--7\%, but, significantly, found that the stars lie on a very tight sequence, with scatter <1--2\%. This, along with their finding of no appreciable correlation between observational indicators of magnetic activity and radius inflation, led them to conclude that stellar magnetism cannot currently explain the radius inflation in main-sequence M dwarfs, a conclusion that dovetails with the above-referenced results suggesting that detached EBs may not be inflated with respect to the single-star population.

The young-age, low-mass region of EB parameter space remains fairly sparsely populated by well-characterised systems, due to the relative scarcity of young, open clusters and the intrinsic faintness of low-mass stars. Therefore, any additional systems with precisely-measured parameters represent benchmark tests of current models. 

Blanco 1 was discovered in 1949 \citep{Blanco1949}. It is an open cluster situated in the local spiral arm, in the direction towards and below the Galactic Centre. The cluster is home to 489 \gaia{} DR2-confirmed member stars at a distance of ${\sim}240$ pc \citep[hereafter \citetalias{GAIA2018}]{GAIA2018}, ranging from A to M spectral types \citep{Gillen2020Rotation}, with $\sim$40 likely brown dwarf members \citep{Moraux2007, Casewell2012}. The metallicity of Blanco 1 is slightly super-solar; \citet{Ford2005} and \citet{Netopil2016} derive [Fe/H] $= +0.04\pm{0.04}$ and [Fe/H] $= +0.03\pm{0.07}$ respectively. It has an on-sky stellar density of ${\sim}30$ stars pc\textsuperscript{-2} \citep{Moraux2007} and a low reddening along the line of sight (E(B-V) ${\sim}0.010$; \citetalias{GAIA2018}). Blanco 1 is in many ways like a smaller, less dense version of the Pleiades (${\sim}110$ Myr, 1326 \gaia{} DR2 members, on-sky stellar density ${\sim}65$ stars pc\textsuperscript{-2}, [Fe/H] ${\sim}-0.01$; \citealt{Moraux2007}, \citetalias{GAIA2018}).

A range of age estimates exists for Blanco 1 from studies made during the past 25 years: $90\pm 25$ Myr based on H$\alpha$ emission \citep{Panagi1997}; $132\pm 24$ Myr and $115\pm 10$ Myr from the lithium depletion boundary (LDB) \citep{Cargile2010, Juarez2014}; $146\pm 14$ Myr based on gyrochronology \citep{Cargile2014}; ${\sim}100$ Myr based on isochrone fitting \citep[hereafter \citetalias{Zhang2020}]{Zhang2020}. Despite the variance, Blanco 1 is known to be young, and the LDB age of 115 $\pm{10}$ Myr \citet{Juarez2014} was found to be a good fit to the lower main sequence of the \gaia{} DR2-confirmed members. Clusters at this age are simultaneously home to low-mass stars contracting down onto the main sequence, intermediate-mass stars in a steady hydrogen-burning state, and high-mass stars evolving off the main sequence \citep{David2016}, making them all the more suited to investigations of stellar evolution.

The Next Generation Transit Survey (\NGTS) survey \citep{Wheatley2017,McCormac2017,2013EPJWC..4713002W,Chazelas2012}, located at ESO's Paranal Observatory, Chile, has been operational since early 2016. Its primary goal is the extension of ground-based transit detections of exoplanets to the Neptune size range, e.g. NGTS-4b, a sub-Neptune-sized planet in the `Neptunian Desert' \citep{West2019}. The enormous amount of data collected by the survey has led to many other interesting discoveries, including the most massive planet orbiting an M-type star NGTS-1b \citep{Bayliss2018}; an ultrashort-period brown dwarf transiting a tidally locked and active M dwarf \citep{Jackman2019}; the most eccentric eclipsing M-dwarf binary system found to date \citep{Acton2020a}; a transiting `warm Saturn' recovered from a \TESS{} single-transit event \citep{Gill2020}; and transit timing variations on the $\sim$540-d-period exoplanet, HIP 41378 f \citep{Bryant2021}. NGTS operations include a survey of nearby open clusters and star forming regions (see \citealt{Gillen2020Rotation} for Paper I and \citealt{Jackman2020} for Paper II), within which the subject of this paper was detected.

We present the identification and characterisation of J00024841-2953539 (hereafter \Nsys) as a triple M-dwarf system in Blanco 1, comprising a short-period EB with tertiary companion. In \S\ref{sec:obs} we describe our observations. We provide details of our modelling procedure in \S\ref{sec:analysis} and give the results in \S\ref{sec:results}. \S\ref{sec:discussion} is a discussion, followed by conclusions in \S\ref{sec:conclusions}.

\section{Observations}
\label{sec:obs}
\Nsys{} was identified as an EB using NGTS photometry. The on-sky separation from its nearest neighbour in the \gaia{} catalogue---a 21-mag object with no parallax measurement---is 24 arcsecs. Objects of comparable brightness are more than 60 arcsecs distant. Follow-up photometry was obtained from SPECULOOS-South \citep{Gillon2018, Burdanov2018, Delrez2018, Murray2020, Sebastian2021} and the South African Astronomical Observatory (SAAO) \citep{Coppejans2013}. It has also been observed by the Transiting Exoplanet Survey Satellite (\TESS) \citep{Ricker2015}. An initial radial velocity (RV) point was taken with HIRES \citep[High Resolution Echelle Spectrometer][]{Vogt1994} on the Keck 10-m telescope in Hawaii. Additional spectra were obtained with UVES (Ultraviolet and Visual Echelle Spectrograph) \citep{Dekker2000},  installed on the VLT (Very Large Telescope) in Paranal (Program ID 0103.C-0902; PI Gillen). The astrometric properties and identifiers for the system are listed in Table \ref{tab:ID_astrometry}. 

\subsection{\NGTS{} photometry}
\label{sub:ngtsphot}
The NGTS facility contains an array of twelve 20 cm wide-field robotic telescopes, each with a 2.8\textdegree{} field of view, 5-arcsec pixels and a 520--890 nm bandpass. The set-up is optimised for observations of K and early M dwarfs. Standard operations, as implemented here, involve 10-second exposures at a cadence of 13 seconds. Aperture photometry is performed with the CASUTools\footnote{http://casu.ast.cam.ac.uk/surveys-projects/software-release} photometry package. Modified versions of the SysRem \citep{Tamuz2005} and Box-fitting least squares \citep{Kovacs2002} algorithms are used for detrending and transit/eclipse detection. Centroiding, as described in \citet{Guenther17b}, is integrated into the pipeline as a means of identifying false positives. Full details of the facility and the reduction pipeline can be found in \citet{Wheatley2017}.

\Nsys{} was identified as an EB in September 2018, following an extended observing campaign on the Blanco 1 open cluster (NGTS field NG0004-2950) \citep{Gillen2020Rotation}. \Nsys{} was observed on 135 nights, producing 201,773 images across 196 days from 2017 May 07 to 2017 November 18, including 45 full eclipses (19 primary and 26 secondary) and numerous partial eclipses.

\begin{table}
	\centering
	\caption{Identifiers and astrometric properties for \Nsys.}
	\begin{tabular}{lll}
	Property	&	Value		&Source\\
	\hline
	\hline
    Identifier	& J00024841-2953539	& 2MASS	\\
    Identifier  & 2320868389659322368 & \gaia{} eDR3\\
    Identifier & TIC 313934158 & \TESS{}\\
    \\
    R.A.		&	\NRA  	& \gaia{} eDR3\\
	Dec			&	\NDec			&  \gaia{} eDR3\\
    $\mu_{{\rm R.A.}}$ (\masy) & \NpropRA &  \gaia{} eDR3\\
	$\mu_{{\rm Dec.}}$ (\masy) & \NpropDec &  \gaia{} eDR3\\
	Parallax (mas)   &   \Nparallax&  \gaia{} eDR3\\
    \hline
    Note	&Epoch is \textbf{J2016.0} for \gaia{} eDR3&\\
    \hline
	\end{tabular}
    \label{tab:ID_astrometry}
\end{table}

\subsection{\textit{TESS} photometry}
\label{sub:tessphot}
\Nsys{} (TIC 313934158) was observed by \TESS\ in Sector 2 between 2018 August 23 and 2018 September 20 (Camera 1; CCD 2). The \TESS\ field of view per camera is $24\times24$ degrees (21 arcsec/pixel) and the bandpass runs from 600 to 1000 nm. We extracted our light curve from the 30-minute cadence full-frame images using the \eleanor{} software package (v1.0.5; \citealt{Feinstein2019}). The eclipse depths are diluted in the \TESS{} light curves due to flux from neighbouring stars, so we tested single-pixel apertures to see how much the effect could be mitigated. In the end, the default two-pixel aperture was selected due to its reduced scatter, because the difference in dilution was minimal and the effect would need to be accounted for in the modelling for any choice of aperture.

\subsection{SPECULOOS photometry}
\label{sub:specphot}
We monitored primary and secondary eclipses using the Callisto telescope at the SPECULOOS-South facility \footnote{https://www.eso.org/public/teles-instr/paranal-observatory/speculoos}, an observatory composed of four semi-robotic independent 1-m telescopes, located at ESO Paranal, Chile. Each telescope is equipped with a deep-depletion 2k\,$\times$\,2k CCD detector optimised for the near infrared, with a $12\times12$ arcmin field of view (0.35 arcsec/pixel). We selected the $I+z'$ filter, which has >90\% transmission from 750 nm to ${\sim}1100$ nm and an exposure time of 60 seconds. Full details of the photometry pipeline can be found in \citet{Murray2020}, but, in brief, the science images are calibrated with standard methods of bias and dark subtraction and flat-field correction, followed by aperture photometry and a differential photometry algorithm, which uses a weighted ensemble of comparison stars to correct for atmospheric and instrumental systematics. The observations took place on the nights of 2018 October 31 and 2018 November 28 for primary and secondary eclipses respectively. 

\subsection{SAAO photometry}
\label{sub:saaophot}
\Nsys{} was observed by the SAAO 1-m telescope on 2020 November 12. Observations were conducted over $\sim$4.5 hrs using the SHOC camera \citep{Coppejans2013}, in the V band. 
The data were bias and flat field corrected via the standard procedure, using the \textsc{SAFPhot}\footnote{https://github.com/apchsh/SAFPhot} Python package. \textsc{SAFPhot} was also used to perform differential photometry on the target, utilising the `SEP' \citep{Barbary2016} package to extract aperture photometry for both the target and nearby comparison stars. The sky background was measured and subtracted by SEP, using a box size and filter width that minimised the background residuals across the frame after the stars had been masked. A 32-pixel box size and 2-pixel box filter were found to give the best results. A single bright comparison star was used to perform differential photometry on the target, with a 2.1 pixel radius aperture, which was found to maximise the signal-to-noise.

\begin{figure}
	\includegraphics[width=\columnwidth]{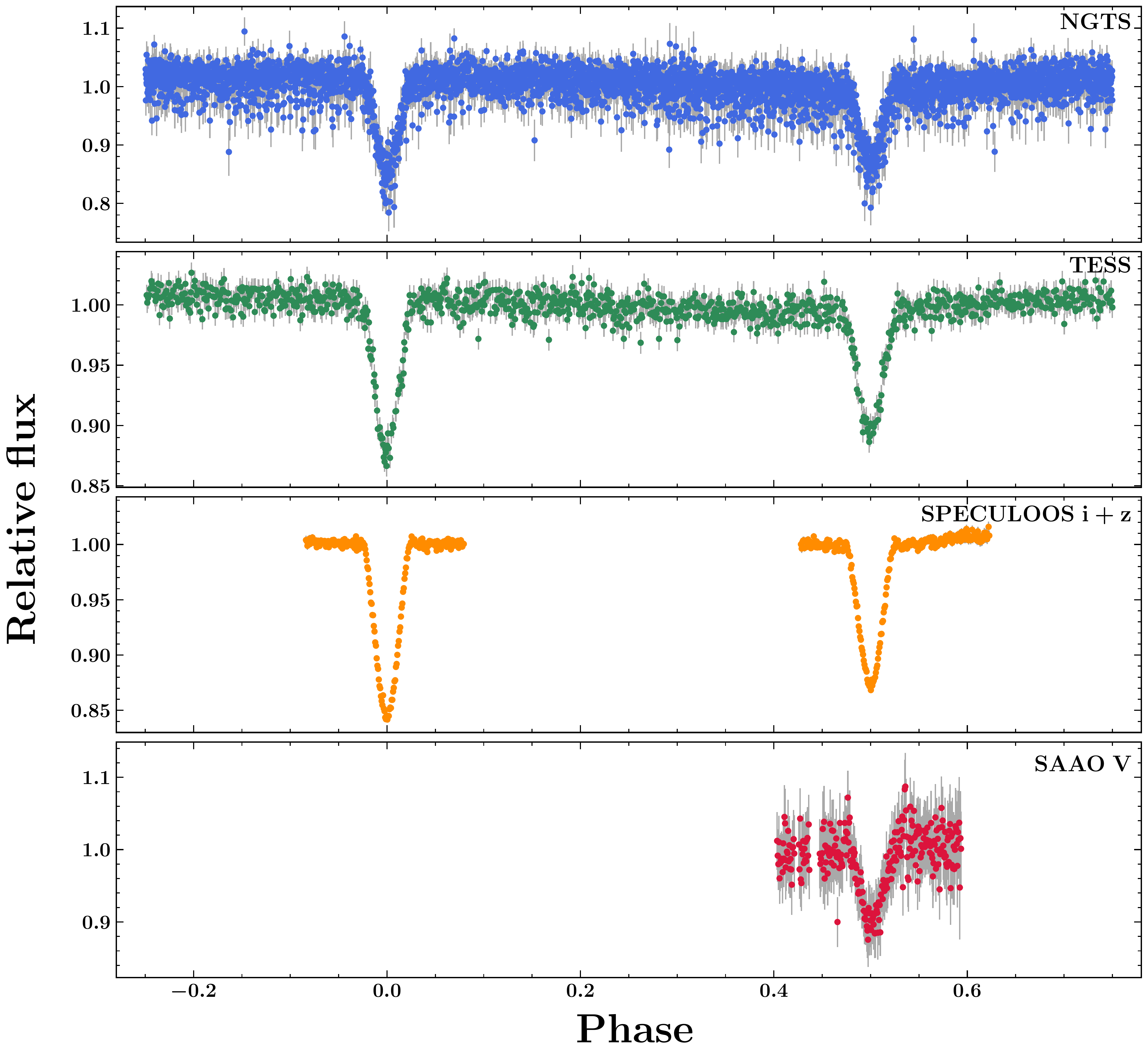}
    \caption{All light curves of \Nsys, normalised and phase-folded on the orbital period. Top to bottom are NGTS, \TESS, SPECULOOS ($I+z'$ band) and SAAO (V band).} 
    \label{fig:allLCs}
\end{figure}

\subsection{HIRES spectroscopy}
\label{sub:HIRESspect}

In order to confirm its nature as a young low-mass cluster member, and spectrally resolve the binary,
the W.M. Keck Observatory Keck I telescope and the facility high-dispersion spectrograph HIRES \citep{Vogt1994}
were used to acquire the first spectrum of the source.  The observation was obtained on 2018 November 11 with the C5 decker,
resulting in a spectrum with resolving power $\mathcal{R}=36\,000$ over 4800--9200 \AA, with some gaps between the redder spectral orders. 
The star Gl 876 (M4 spectral type) was observed as a radial velocity standard.
The two-dimensional spectral images were processed with the MAKEE pipeline reduction package, written by Tom Barlow. These observations revealed that the eclipsing system was indeed a triple.

\subsection{UVES spectroscopy}
\label{sub:UVESspect}
We obtained multi-epoch spectroscopy for \Nsys{} between June and August 2019 with the red arm of UVES. The spectra were exposed onto a mosaic of two 2k\,$\times$\,4k CCDs (EEV + MIT/LL) with 15 µm pixels and a pixel scale of 0.182 arcsec/pixel. The resulting usable wavelength coverage extended from 6700 to 9850 \AA{} with a gap of ${\sim}100$ \AA{} at the centre. The observations were taken using a spectrograph slit width of 1.2 arcsec with resolving power $\mathcal{R}\sim40\,000$. We opted for 2$\times$2 on-chip binning, giving median signal-to-noise (measured at the order centres for all orders and epochs) of 16.9 and 25.2 for the two respective CCDs. The data were reduced with the standard UVES pipeline recipes \citep{Ballester2000} (version 5.10.4) via EsoReflex \citep{Freudling2013}, and we made use of both the individual echelle orders and the merged spectra created. We downloaded raw data of the RV standard star GJ 109 (M3V spectral type, \citealt{Henry2002}) from the UVES archive (Program ID 074.B-0639(A) and $\mathcal{R}\sim46\,000$) to be used as a template spectrum. The same reduction methods were implemented, producing spectra with median signal-to-noise of 284 and 217 for the two CCDs. GJ 109 appears in \cite{Nidever2002} with velocity scatter below 0.1 km s\textsuperscript{-1}. It is also one of the validation stars used in the construction of the \gaia{} catalogue of RV standard stars \citep{Soubiran2018}, wherein 47 RVs collated from the SOPHIE spectrograph \citep{Perruchot2008} show a standard deviation of 0.013 km s\textsuperscript{-1} over a 12-year baseline. Extending the work of \citet{Nidever2002}, \citet{Chubak2012} added to the catalogue of high precision Keck-HIRES RV measurements for FGKM stars. For this work, we adopt the mean barycentric RV from \citet{Chubak2012} of 30.458 km s\textsuperscript{-1}. The listed standard deviation is 0.149 km s\textsuperscript{-1} over 11 measurements. This uncertainty, the standard error of the mean, or the formal uncertainties in the other catalogues, are small compared with the known systematic uncertainties inherent to M-dwarf RVs, e.g. due to gravitational redshift and convective blueshift. This uncertainty is ${\sim}$0.3 km s\textsuperscript{-1} \citep{Kraus2011, Chubak2012}, which we adopt as the contribution from our template.

\begin{table*}
	\centering
	\caption{Radial Velocities for \Nsys.}
	\label{tab:rvs}
	\begin{tabular}{cccccccc} 
		\hline \hline
\multicolumn{3}{c}{Epoch} & S/N* & \multicolumn{3}{c}{RV [km/s]} & Instrument\\
		\cline{1-3} \cline{5-7}
    \noalign{\smallskip} 
UT date & BJD\textsubscript{TDB} &	Phase &   & Primary & Secondary  & Tertiary\\
		\hline
2018-11-03&2458425.90890&	0.116&  19 &	$-33.13 \pm 1.02$&	$81.72 \pm 2.27$&$10.42 \pm 2.14$ &HIRES\\
2019-06-14&2458648.88849&	0.193&	21&	$-54.85 \pm 0.53$&  $109.79 \pm 0.80$&  $5.28 \pm 1.24$&UVES\\
2019-07-06&2458670.86731&	0.210&	21&	$-55.81 \pm 0.55$&	$114.10 \pm 0.89$&  $3.31 \pm 1.22$&UVES\\
2019-07-10&2458674.77115&	0.765&	21&	$67.94 \pm 0.49$&	$-106.36 \pm 0.73$&  $4.97 \pm 1.18$&UVES\\
2019-07-10&2458674.89594&	0.879&	21&	$49.16 \pm 0.42$&	$-72.60 \pm 0.71$&  $4.24 \pm 0.86$&UVES\\
2019-07-12&2458676.82505&	0.636&	20&	$53.36 \pm 0.54$&	$-79.25 \pm 0.91$&  $5.90 \pm 1.23$&UVES\\
2019-08-02&2458697.76597&   0.708&	17&	$66.84 \pm 0.51$&	$-102.52 \pm 0.77$&  $7.21 \pm 1.27$&UVES\\
2019-08-04&2458699.89145&   0.643&	21&	$55.06 \pm 0.50$&	$-83.00 \pm 0.84$&  $6.19 \pm 1.06$&UVES\\
2019-08-10&2458705.87215&  0.090&	14&	$-27.32 \pm 0.55$&	$64.96 \pm 0.81$&  $7.49 \pm 1.24$&UVES\\
2019-08-10&2458705.88096&  0.098&	13&	$-30.49 \pm 0.66$&	$70.53 \pm 1.04$&  $7.96 \pm 1.58$&UVES\\
		\hline
	\end{tabular}

    \begin{tabular}{l}
    *Note: HIRES S/N: at 7040 \AA{} continuum. UVES S/N: median values based on orders used in the RV extraction ($\sim$7800 \AA).\\
    \end{tabular}
\end{table*}


\section{Analysis}
\label{sec:analysis}

\begin{figure*}
	\includegraphics[width=\textwidth,angle=0]{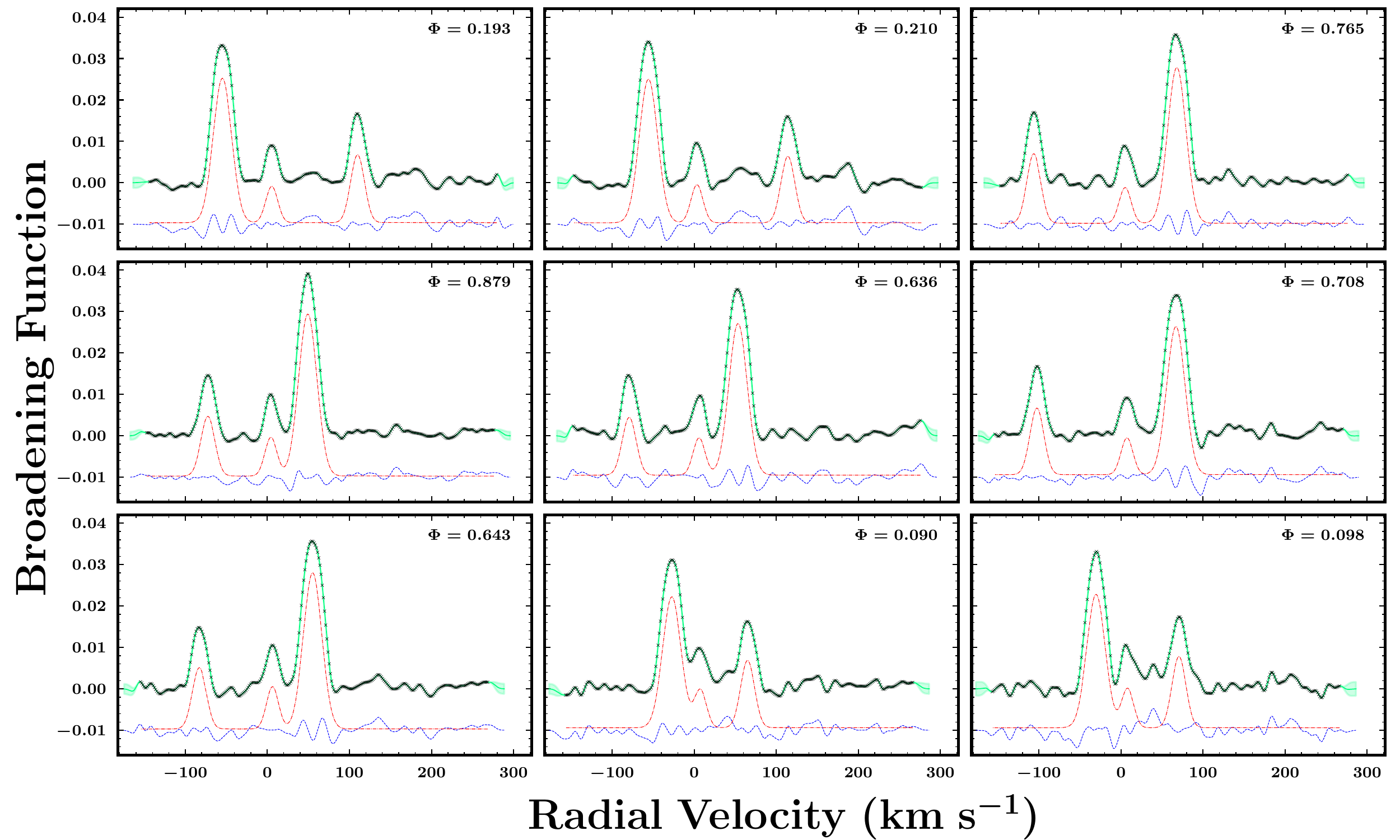}
    \caption{Broadening Functions (black crosses) from our UVES spectra at each epoch in observation order, with the orbital phases labelled. Each plot shows the best-fit model (green) together with the associated 68.3\% confidence interval (light green shaded region). The Gaussian (red dot-dashed line) and GP (blue dashed line) terms are also displayed, with a vertical offset added for clarity. The tallest peak in each plot corresponds to the primary component, whilst the second and third tallest peaks correspond to the secondary and tertiary components respectively.}
    \label{fig:BFexample}
\end{figure*}

\subsection{Radial velocities}
\label{sub:rvs}

For the single HIRES observation, a traditional cross correlation technique was used to reveal three spectral components
in the system.  As in \cite{Gillen2017,Gillen2020EB}, RVs were derived 
using the {\it fxcor} package within IRAF \citep{Tody1986} 
to correlate the spectrum of \Nsys{} with that of the spectral standard observed on the same night,
plus three other spectra of this same standard taken on different nights.  This was done in order to improve the error  
on the measurement, since the night-to-night differences in the derived RVs are smaller than the order-to-order differences.
Regions of telluric contamination were avoided within the wavelength range 6330--7160 \AA{} used in deriving the average velocities
for the three components that are reported in Table \ref{tab:rvs}. The velocities are the medians, while the uncertainties are the standard deviations
among all standard spectra and orders.

We extracted the UVES RVs using the Broadening Function (BF) technique as introduced by \cite{Rucinski1992, Rucinski1999, Rucinski2002}\footnote{Also introduced by \citet{Rix1992} in the context of line-of-sight velocities of galaxies and applied to Doppler imaging of starspots by \citet{Barnes2004}.}. The approach uses singular value decomposition (SVD) to determine the Doppler broadening kernel, $B$, from its assumed convolution with a template spectrum, $T$, when we observe target spectrum $S$:
\begin{equation}
S = B * T. 
\end{equation}

This method holds some advantages over the cross correlation technique when analysing rotationally-broadened spectra of binary or higher-order systems. For example, whilst the cross-correlation function (CCF) inherits the common broadening components of both template and target star, such as instrumental, thermal and micro-turbulence \citep{Rucinski1999}, the BF isolates the rotational broadening contribution (assuming the main difference between the target and template stars' spectra is attributable to rotation), and so offers superior resolution, as well as being less susceptible to the ``peak pulling'' effect, which can be an issue when peaks overlap \citep{Rucinski2002, Hensberge2007, Kraus2011}. 

Rucinski provides IDL routines and a description of the method\footnote{http://www.astro.utoronto.ca/~rucinski}, which we used as a basis for our own \Python{} implementation. We refer the reader to the above (and to \cite{Rucinski1992, Rucinski1999, Rucinski2002}) for a detailed description, but we give a brief summary of the process here. The UVES observation times were converted to BJD (Barycentric Julian Date) format in the TDB (Barycentric Dynamical Time) time system, and barycentric corrections were computed. Each spectrum was continuum normalised using a cubic spline, with outliers ($+5$ sigma and $-8$ sigma) removed using a rolling median filter. The spectra were re-sampled to a common wavelength vector in equal increments of log lambda with slight over-sampling. Regions significantly affected by telluric lines were removed, before the broadening functions, with 301 bins, were computed with the SVD module in \PyAstronomy{} \citep{pya} and smoothed with a Gaussian.

In the wavelength regime of our observations (6700--9850 \AA{}) there is significant atmospheric absorption due to water and oxygen. Consequently, many echelle orders were unsuitable for deriving RVs. Based on Cerro Paranal's yearly precipitable water vapour \citep{Moehler2014} and ESO's Sky Model \citep{Noll2012, Jones2013}, and assuming airmass = 1.0 (closest available to our observations), we identified the wavelength regions in our UVES observations where atmospheric transmission was expected to be better than 95\%. With an additional criterion of a 50 \AA{} minimum length, the resulting regions were contained in 12 echelle orders, spanning approximately 6700 to 8880 \AA. 

The UVES pipeline merges echelle orders into a single 1D spectrum; one for the lower CCD and one for the upper CCD. We trialled approaches using either individual orders or the merged spectrum, with very similar results, but we report values based on the merged spectrum, where the BFs were slightly better defined, yielding more precise RVs. This method took the merged spectrum from the lower CCD for each epoch and removed the telluric regions, leaving a single spectrum per epoch from which BFs were calculated. The usable spectral window was slightly smaller in this case (6750--8110 \AA) due to the fact that the upper CCD was not incorporated. The segments of the merged spectrum used were as follows: 6740--6866 \AA, 7055--7150 \AA, 7386--7560 \AA, 7713--7894 \AA{} and 8031--8110 \AA. Whilst this approach relies on sensible merging of orders in the UVES pipeline, it benefits from the target spectrum being significantly longer than the BF, which is advantageous because the quality of the determination of the BF increases in relation to how many times the spectrum is longer than the BF \citep{Rucinski2002}. 

A certain level of noise is invariably present along the baseline of BFs and CCFs. This underlying structure is worthwhile accounting for, because it can subtly affect the peak shapes and apparent centres from which the RVs are measured. To that end, and to assist in making robust uncertainty estimates, we chose to model the noise in the BFs with a Gaussian process (GP) at the same time as fitting for the peaks, an approach applied to CCFs in \cite{Gillen2014}. Three clear peaks, suggestive of a triple system, are present in all spectra. Therefore, each BF was modelled as the sum of three Gaussians with a small constant offset, plus a GP with squared exponential kernel (see Figure \ref{fig:BFexample}). All BFs were modelled simultaneously with 13 parameters fit to each: height, width and location of each Gaussian, vertical offset, two GP hyperparameters and a small white noise term. The Gaussian profile widths were deemed to be constant between all BFs and so were jointly fit. The posterior parameter space was explored using the affine invariant Markov Chain Monte Carlo (MCMC) method implemented in \emcee{} \citep{foremanmackey13}, with 500 `walkers'. The parameter values were initialised around estimates from a simple three-Gaussian fit and some trial runs. The chains were run for 200\,000 steps, the first 100\,000 steps were discarded as `burn-in', and each chain was thinned based on the average auto-correlation time. The GP component was handled with the \george{} package \citep{Ambikasaran2015}. In Table \ref{tab:rvs}, we report the median samples from the relevant marginalised distributions as the UVES RVs. Our adopted uncertainty for each RV is the mean of the values corresponding to the 16th and 84th percentiles (which are always consistent at the 1--2\% level), plus the estimated uncertainty for the template star of 0.3 km s\textsuperscript{-1}, added in quadrature. In the global model, the UVES and HIRES RVs were modelled using Keplerian orbits. A jitter term was fitted to the UVES RVs, along with an offset term for the single HIRES observation in order to account for the difference between instrument zero points.

The broadening function methodology was additionally applied to the derivation of spectroscopic light ratios from the UVES spectra, which were subsequently used as constraints in the global modelling. These light ratios were determined by measuring the areas under the Gaussian peaks fitted to the BFs. This was done for three wavelength segments across the UVES spectra: 6740--7150 \AA{}, 7386--8110 \AA{} and 8690--8882 \AA{}. For each segment, we computed BFs for all epochs and fitted them with the GP + three-Gaussian model described previously. We then measured the area under each Gaussian peak and computed the final light ratios by taking an inverse-variance weighted mean for each wavelength region, propagating the uncertainties through from the MCMC fit. The values determined for the binary and tertiary light ratios are shown in Table \ref{tab:light_ratios}, where the quoted uncertainties are the standard errors of the weighted means, $\big(\sqrt{\sum_{i=1}^{n}w_{i}}\big)^{-1}$.

\begin{table}
	\centering
	\caption{Binary and tertiary light ratios from UVES spectra.}
	\begin{tabular}{ccc}
	\hline
	\hline
	Wavelength region	&	Binary light ratio		&Tertiary light ratio\\
	\AA &$l_{\textrm{sec}}/l_{\textrm{pri}}$&$l_{\textrm{ter}}/(l_{\textrm{pri}}+l_{\textrm{sec}}+l_{\textrm{ter}})$	\\
	\hline
     6740--7150 & $0.30\pm{0.01}$	& $0.10\pm{0.01}$\\
     7386--8110  & $0.37\pm{0.01}$	&$0.15\pm{0.01}$\\
     8690--8882  &	$0.38\pm{0.02}$	& $0.15\pm{0.01}$\\
    \hline
	\end{tabular}
    \label{tab:light_ratios}
\end{table}

\subsection{Global modelling}
\label{sub:Global_modelling}
The global modelling was performed with GP-EBOP, an eclipsing binary and transiting planet model that is optimised for modelling young and/or active systems. We give a brief description here, but refer readers to \citet{Gillen2017} and \citet{Gillen2020EB} for more details.

GP-EBOP permits simultaneous modelling of light curves, RVs and SEDs, using a GP framework to model the out-of-eclipse (OOE) variations and stellar activity. The GP model means that uncertainties in the variability modelling can be propagated through to the posterior distributions for the EB parameters. GP-EBOP uses an EB model based on that described in \citet{Irwin2011, Irwin2018}, which is a descendent of the EBOP family of models, but which uses the analytic method of \citet{Mandel2002} to perform the eclipse calculations. Limb darkening is parameterised using the triangular sampling method of \citet{Kipping2013}, with theoretical constraints applied based on the predictions of the Limb Darkening Toolkit (LDtk; \citealt{Parviainen2015}). The posterior parameter space is explored using \emcee.

There are two main updates to GP-EBOP since \citet{Gillen2020EB}:
\begin{enumerate}
\item The GP model can now optionally use the \celeritetwo{} library \citep{celerite2}, as well as \celerite{} \citep{celerite1} and \george.
\item GP-EBOP is able to simultaneously model the component SEDs of triple systems.
\end{enumerate}

For the analysis presented here, we simultaneously modelled the observed light curves, RVs and SED of \Nsys{}.

\subsection{Light curves}
\label{sub:LCs}
The NGTS light curve (see Figure \ref{fig:allLCs}) was sigma clipped outside of the eclipses on a nightly basis with a 5-sigma ($\textrm{sigma} = 1.4826~\times~\textrm{Median Absolute Deviation (MAD)}$) threshold. The time array was converted to BJD\textsubscript{TDB} with \astropy{} \citep{astropy:2013,astropy:2018} and then the light curve was median-normalised and binned in time to 10 minutes. An additional 5-sigma nightly clipping was applied to the binned light curve, as well as the removal of two nights badly affected by adverse observing conditions. As is evident from Figure \ref{fig:allLCs}, the system displays a gentle modulation in flux on the orbital period, which peaks around primary eclipse. The peak-to-trough variation is 1.8\% and 1.2\% in the NGTS and \tess{} light curves respectively. The most-likely cause of this modulation is starspots, which explains the difference in amplitudes between NGTS and \tess{} (the \tess{} passband being redder). The low-amplitude signal suggests that any longitudinal inhomogeneities on the stellar surfaces are modest. We also find a little variability in phase with the lunar cycle, a consequence of imperfect background subtraction known to affect fainter (NGTS $\gtrsim15$ mag) targets. For the GP component of the model, we chose the rotation kernel implemented in \celeritetwo, which is a good descriptive model for a wide range of stochastic variability in stellar time series, including rotational modulation. The kernel is a mixture of two stochastically-driven, damped harmonic oscillator (SHO) terms. The power spectral density of the SHO term is given by
\begin{equation}
S(\omega) = \sqrt{\frac{2}{\pi}} \frac{S_{0}\,{\omega_{0}}^2}
{{({\omega}^2 - {\omega_{0}}^2)}^2 + {\omega_{0}}^2\,{\omega}^2 / {Q}^2},
\end{equation}
where $\omega$ is the angular frequency, $S_{0}$ is the amplitude of the oscillation, $\omega_{0}$ is the un-damped frequency, and $Q$ is the quality factor. We checked that the resulting interpolations across each eclipse were satisfactory, with the eclipse depths maintained before and after detrending for the GP component. We initialised the MCMC sampler using the well-determined ephemeris, with wide uniform priors placed on all GP hyperparameters. 

The GP kernel adopted for the \TESS{}, SPECULOOS and SAAO light curves was a single SHO term with $Q = 1/\sqrt{2}$. The \TESS{} light curve was sigma clipped outside of the eclipses using a median filter and a 10-sigma threshold, which removed two outlier points. We considered the uncertainties from \eleanor{} to be too large given the scatter, and so calculated our own for each data point, \textit{i}, as
\begin{equation}
\sigma_{i,\textrm{new}}=1.4826 \times \textrm{MAD}_\textrm{OOE} \times \frac{\sigma_{i,\textrm{old}}}{\textrm{median}(\sigma_\textrm{old})}.
\end{equation}

The high-precision SPECULOOS observations contain single primary and secondary eclipses. The data displayed ramps in flux at the start of each night, which are characteristic of ground-based observations looking through high airmass. Accordingly, the first six data points taken in each night were removed before modelling.

The SAAO light curve was sigma clipped using a median filter and a 5-sigma threshold. Diagnostic plots from the observations revealed movement across the CCD, as well as variations in the background flux taking place during eclipse. We attribute the slight asymmetry in the eclipse shape to these effects. We also note that the inclusion of the SAAO light curve led to greater uncertainty in the derived radius of the secondary. This could be explained by the fact that the noise and asymmetry found in the SAAO secondary eclipse is not present in any of the other light curves, and so the global model does not significantly change to account for these features. The consequent poorer fit to the SAAO light curve has the effect, however, of increasing the uncertainties. Our intention with the V-band SAAO observations was to have a constraint in a bluer spectral region, which might break the degeneracy between radius ratio, inclination and surface brightness, thought to be the cause of the poorly constrained radius ratio we obtained when modelling the light curves and RVs without the SED. However, bad weather thwarted attempts to obtain a primary eclipse with SAAO, leaving this constraint unrealised. We include the secondary eclipse in our global model nonetheless.

\subsubsection{Gravity darkening}
\label{sub:grav_dark}
We adopted gravity darkening coefficients from the tables of \citet{Claret2011} for log\,$g=5.0$, solar metallicity, \textit{I}-band filter and PHOENIX atmosphere models, interpolating to the temperatures of each binary component. These tables give the required input to the central EB model \citep{Irwin2018}.

\subsection{Spectral energy distribution}
\label{sub:sed}
\begin{table*}
	\centering
	\caption{Broadband photometry constituting the observed SED of \Nsys{}.}
	\label{broadband}
	\begin{tabular}{lllll} %
	\hline
	\hline
	Band	& System &	Magnitude		&Spectral flux density &Refs.\\
	       &         &   &$(\mathrm{erg~s^{-1}~cm^{-2}}$ \AA$^{-1})$ &\\
	\hline
    Pan-STARRS1 \textit{g}&AB  &\PANgmag &\PANgfl &(1,7,8)\\
    Pan-STARRS1 \textit{r}&AB  &\PANrmag &\PANrfl &(1,7,8)\\
    Pan-STARRS1 \textit{i}&AB  &\PANimag &\PANifl &(1,7,8)\\
    Pan-STARRS1 \textit{z}&AB &\PANzmag &\PANzfl &(1,7,8)\\
    Pan-STARRS1 \textit{y}&AB &\PANymag &\PANyfl &(1,7,8)\\
    APASS \textit{g}&AB  &\APgmag &\APgfl &(2,7)\\
    APASS \textit{r}&AB  &\APrmag &\APrfl &(2,7)\\
    APASS \textit{i}&AB  &\APimag &\APifl &(2,7)\\ 
    SkyMapper \textit{g}&AB &\SKYgmag &\SKYgfl &(3,7)\\
    SkyMapper \textit{r}&AB &\SKYrmag &\SKYrfl&(3,7)\\
    SkyMapper \textit{i}&AB &\SKYimag &\SKYifl&(3,7)\\
    SkyMapper \textit{z}&AB  &\SKYzmag &\SKYzfl&(3,7)\\
    \gaia{} G&Vega		&\NGAIAGmag	&\NGAIAGfl &(4,7,9)\\
    \gaia{} G\textsubscript{BP}&Vega	&\NGAIABmag	&\NGAIABfl &(4,7,9)\\
    \gaia{} G\textsubscript{RP}&Vega   &\NGAIARmag	&\NGAIARfl&(4,7,9)\\
    2MASS \textit{J}&Vega		&\NJmag		&\NJfl&(5,7,10)\\
   	2MASS \textit{H}&Vega 	&\NHmag		&\NHfl&(5,7,10)\\
	2MASS K\textsubscript{s}&Vega 	&\NKmag		&\NKfl&(5,7,10)\\
    WISE W1&Vega 	&\NWmag		&\NWfl &(6,7)	\\
    WISE W2&Vega 	&\NWWmag	&\NWWfl	&(6,7)	\\

	\hline
	\end{tabular}
	\\ 
	\begin{flushleft}
	\textbf{References.} \textit{Photometry:} 1. \citet{Chambers2016}; 2. \citet{Henden2019}; 3. \citet{Onken2019}; 4. \citet{GaiaDR3_summary}; 5. \citet{2MASS}; 6. \citet{WISE};. \textit{Bandpasses:} 7. Filter Profile Service (FPS: \url{http://svo2.cab.inta-csic.es/theory/fps}); 8. \citet{Tonry2012}; 9. \citet{GaiaDR3photometry}; 10. \citet{Cohen2003};. 
	\end{flushleft}

    \label{tab:broadband}
\end{table*}

Table \ref{broadband} reports our collected broadband photometric measurements which we used to model the SED of \Nsys{}. This photometry covers the rise, peak and fall of the combined stellar photospheric emission spectra, thus providing useful constraints on the effective temperatures. 

We modelled the observed SED of \Nsys{} as the sum of three stellar photospheres (primary, secondary and tertiary components). Model grids of BT-Settl model atmospheres \citep{Allard2012} were convolved with the spectral response functions of each band to create a grid of model fluxes in steps of 100 K in effective temperature and 0.5 in surface gravity over ranges $1200\leq T_{\textrm{eff}} \leq 7000$ and $3.0\leq \textrm{log}\,g \leq 5.5$. Each SED was modelled by interpolating the model grids with a cubic spline in $T_{\textrm{eff}}$--$\textrm{log}\,g$ space, keeping metallicity fixed at $Z=0.0$. The observed magnitudes were converted to spectral flux densities using standard relations, with the zero-point values and effective wavelengths obtained from the references in Table \ref{broadband}. The parameters of the fit were: the temperatures, radii and surface gravities of both stars, the distance and reddening to the system, and a jitter term per photometric dataset. The reddening model follows the extinction law of \citet{Fitzpatrick1999} with the improvements made by \citet{Indebetouw2005}.

Simultaneously modelling the SED, along with the light curves and RVs, enables measurements of the spectroscopic light ratios (as described in \S\ref{sub:rvs}) to constrain the eclipse modelling in the observed light curve bands. It also means that we solve for the stellar masses, radii and temperatures in a self-consistent manner. 
We implemented the spectroscopic light ratios by applying prior constraints on the model binary ($l_{\textrm{sec}}/l_{\textrm{pri}}$) and tertiary ($l_{\textrm{ter}}/(l_{\textrm{pri}}+l_{\textrm{sec}}+l_{\textrm{ter}})$) light ratios within the UVES band. Those constraints were propagated into the eclipse modelling by using the corresponding atmospheric model ratios of emergent stellar fluxes and luminosities in the NGTS, \TESS{}, SPECULOOS and SAAO bands as the central surface brightness ratios and third light parameters given to the central EB model. In the case of \TESS{}, we fitted for an extra third light component (added to the model tertiary light ratio), due to the eclipse dilution previously described. We allowed the light ratio uncertainties to inflate by fitting jitter terms to each. These were added to account for any additional uncertainties in the measurement procedure, e.g. from the use of broadening functions and Gaussian fits; the use of a single spectroscopic template which cannot be a perfect match to all three stellar components; and the use of stellar atmosphere models. A modified Jeffreys prior was placed on each jitter term, with the `knee' value set to twice the calculated uncertainty of the corresponding light ratio, and the upper bound set to 0.5. The propagation of these spectroscopic constraints helped to break the degeneracy between radius ratio, inclination and surface brightness. 

The use of light ratios derived from spectral lines as proxies for broadband flux ratios is reasonable when the stars in question have similar spectral characteristics, but it can pose potential problems in other cases. That is, for stars with significantly different spectral types, and hence differing line strengths, spectroscopic light ratios may be inadequate representations of passband-integrated light ratios. For the present case of three mid--late M dwarfs, where we use wide spectral windows for the calculation of the BFs, which should help in averaging out any differences in particular spectral lines, we expect the approach to be valid.

The use of stellar atmosphere models to predict flux ratios between stars in different photometric bands means that a model dependence is introduced. Whilst this is not an ideal approach for the derivation of EB parameters, we expect the model dependence to be small, because while atmosphere models do not reproduce all spectral lines and features, they should be able to provide reasonable constraints on the flux ratios between two or three model atmospheres in wide photometric bands, such as those of our observations and broadband photometry. We note that a test case for this method appeared in \citet{Gillen2020EB}, where a comparison was made---for a system without significant degeneracies in radius ratio, inclination and surface brightness---between modelling the light curves and RVs only, and also modelling the SED; consistent masses and radii were found. We also note that, of the fundamental parameters, it is only the radius that is in any meaningful way subject to additional model dependence compared with standard EB parameter derivation, e.g. the masses are almost entirely constrained by the RVs, and individual effective temperatures are always reliant on theoretical and/or empirical relations, or else SED modelling as performed here.

\begin{figure*}
	\includegraphics[width=0.96\textwidth]{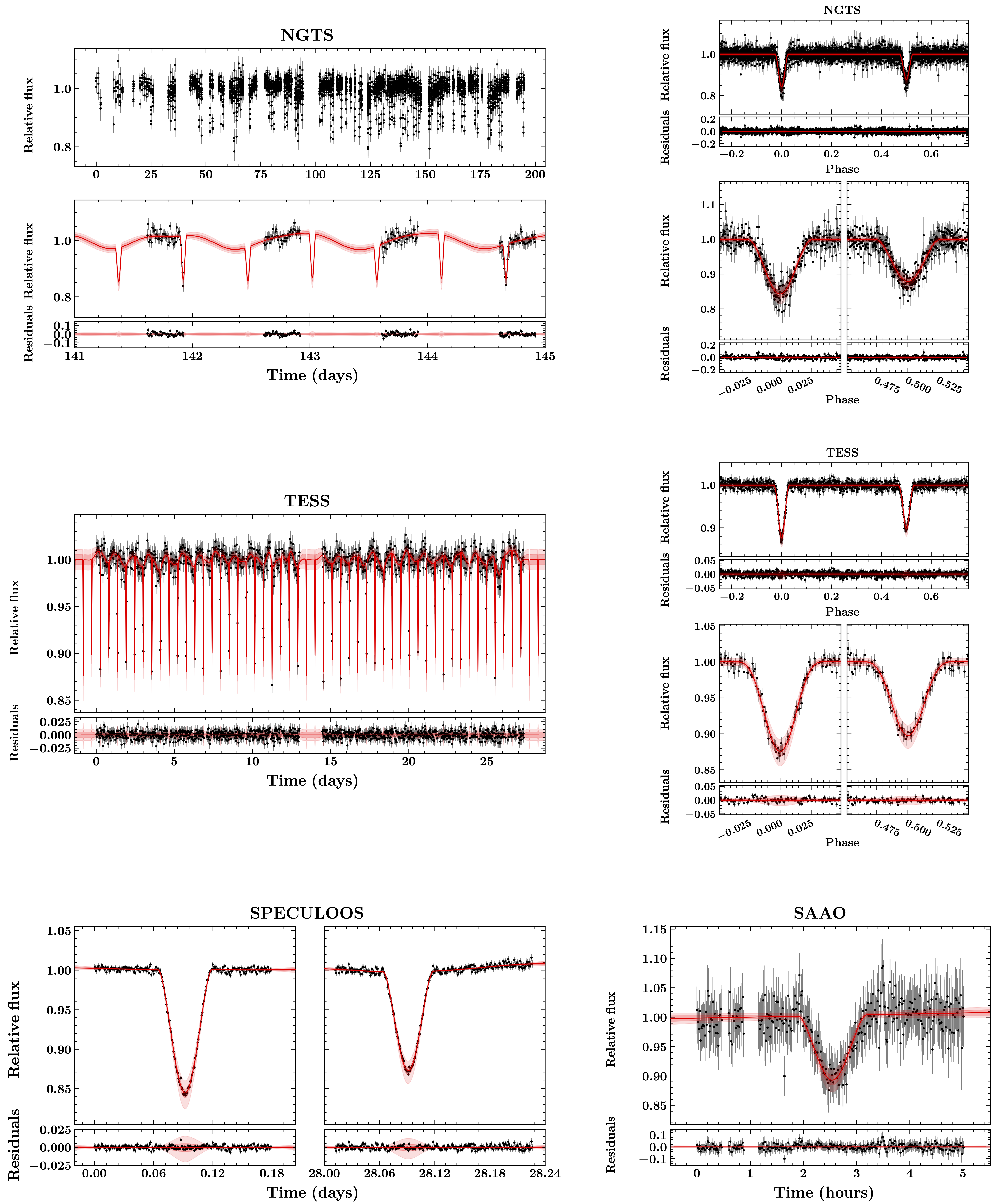}
    \caption{Top left: NGTS relative flux light curve of \Nsys{} (binned to 10 mins), with a 4-day close-up in the lower panel showing the GP-EBOP model in red and residuals below. The red line and pink shaded regions show the mean and 1- and 2-sigma confidence intervals of the predictive posterior distribution. Centre left: As above, but for the \tess{} observations. In this case the GP model is over-layed on the whole light curve. Bottom left and right show equivalent plots for SPECULOOS and SAAO light curves respectively. Top and centre right: phase-folded light curves from NGTS and \tess{} respectively, which have been detrended with respect to the Gaussian process model. The red line indicates the median EB model derived from the posterior distribution, i.e., individual draws are calculated across phase space and the median of their paths plotted. Phase zero marks the center of the primary eclipse. Immediately below are the residuals of the fit. The lower panels display zooms on primary and secondary eclipses (left and right, respectively) with the median model and 1- and 2-sigma uncertainties shown (red line and pink shaded regions, respectively). Residuals are shown immediately below.}
    \label{fig:LCs_GPEBOP}
\end{figure*}

\begin{figure}
	\includegraphics[width=\columnwidth]{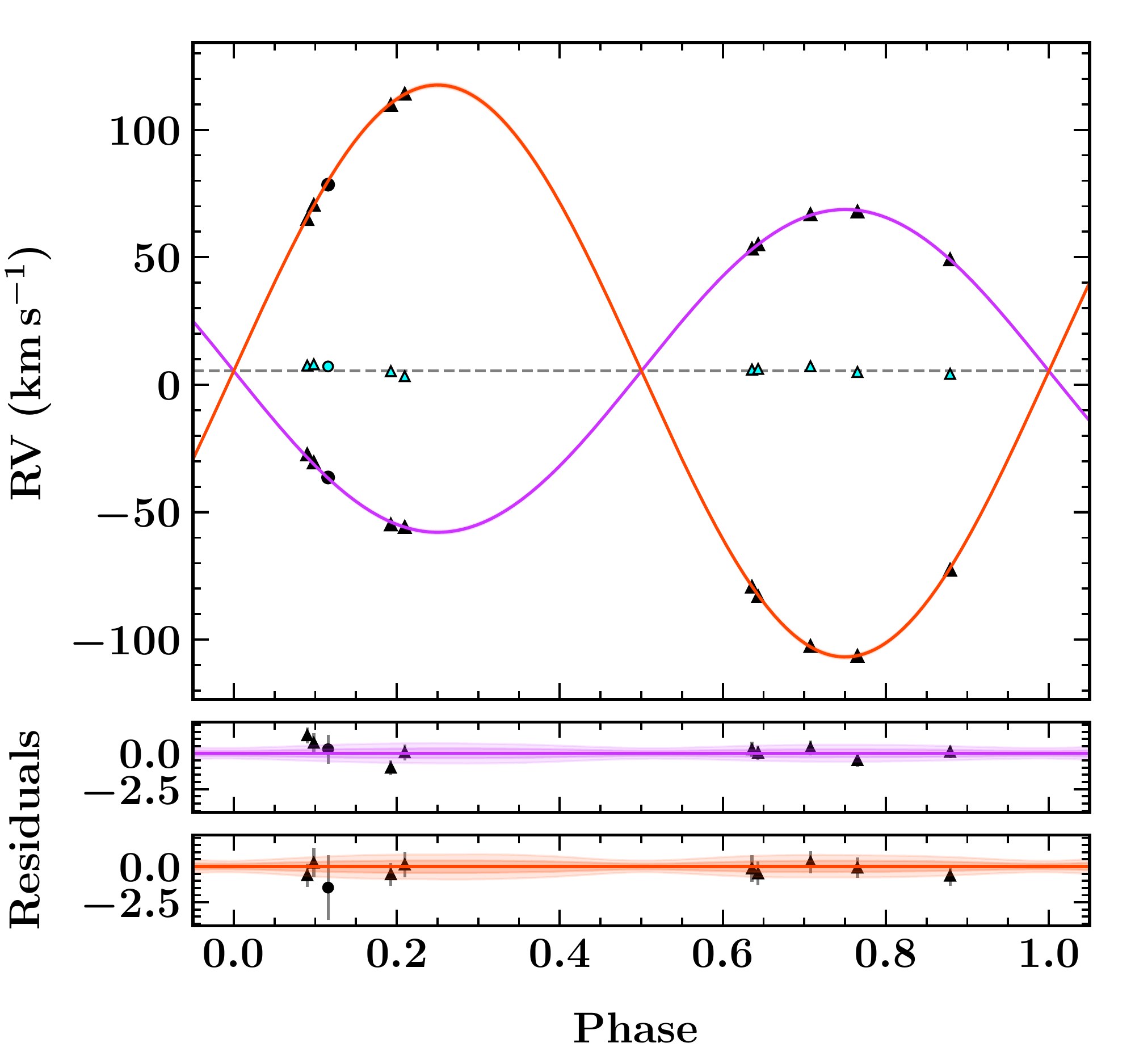}
    \caption{Top: phase-folded RV orbit of \Nsys{} with UVES (triangles) and HIRES (circles) RV measurements for the primary and secondary stars (purple and orange, respectively). The lines and shaded regions indicate the median and 2$\sigma$ uncertainty on the posterior distribution of the RV orbits. The tertiary component RVs are plotted with cyan markers and the grey horizontal dotted line indicates the systemic velocity. Bottom: residuals of the fit.}
    \label{fig:rvs}
\end{figure}

\begin{figure}
	\includegraphics[width=\columnwidth]{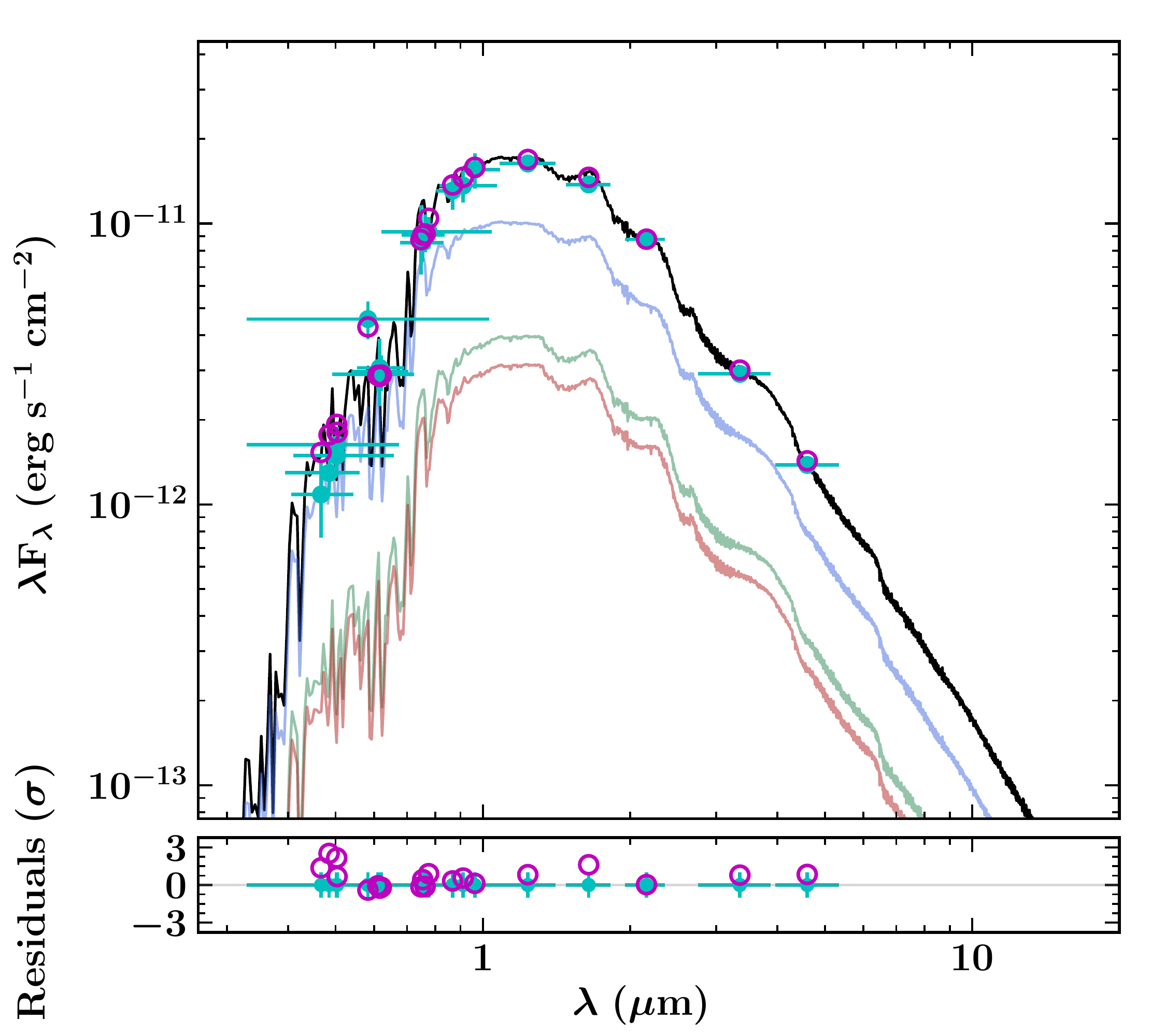}
    \caption{Spectral energy distribution of \Nsys{}. Cyan points represent the observed broadband magnitudes reported in Table~\ref{tab:broadband}, which together comprise the observed SED. The horizontal cyan error bars indicate the spectral range of each band. SEDs constructed from BT-Settl model atmospheres for the primary, secondary and tertiary stars are shown in blue, green and red respectively. Their combined SED is shown in black, with its prediction in each observed band indicated by magenta circles. Residuals are shown below.}
    \label{fig:sed}
\end{figure}

{\renewcommand{\arraystretch}{1.3}
\begin{table*}  
 \centering  
 \caption[Model parameters]{Fitted and derived parameters for \Nsys{}.}  
 \label{tab:params}  
 \begin{tabular}{l l c l }  
 \noalign{\smallskip} \hline  \hline \noalign{\smallskip}  
 Parameter  &   Symbol  &  Value  &  Unit \\   
 \noalign{\smallskip} \hline 
\multicolumn{4}{c}{\emph{Fitted physical parameters}} \\   
\hline  \noalign{\smallskip}

 Orbital period    &    $P$    &  \period  &    days      \\
 Time of primary eclipse centre    &    $T_{\rm{prim}}$    &  \epoch   &  BJD \\
 Sum of radii    &    $(R_{\rm{pri}} + R_{\rm{sec}})/ a$    &  \rsum   &   \\
 Radius ratio    &    $R_{\rm{sec}} / R_{\rm{pri}}$    &  \rratio   &   \\
 Cosine of orbital inclination & $\cos i$ &   \cosi  &     \\
 Eccentricity and argument of-    &   $\sqrt{e} \cos \omega$   & \recosw  &  \\
 -periastron combination terms    &   $\sqrt{e} \sin \omega$   &  \resinw   &   \\
 Systemic velocity    &    $V_{\rm{sys}}$    &   \velsys   &  km\,s$^{-1}$  \\
 Primary RV semi-amplitude    &    $K_{\rm{pri}}$    &  \rvpri  &  km\,s$^{-1}$   \\
 Secondary RV semi-amplitude    &    $K_{\rm{sec}}$    &  \rvsec  &  km\,s$^{-1}$   \\
 
 Distance    &    $d$    & \Dist  &  pc    \\
 Reddening    &    A$_{\rm V}$  &  \Av  &      \\
 Primary effective temperature    &    $T_{\rm pri}$    &  \Tpri &  K   \\
 Secondary effective temperature    &    $T_{\rm sec}$    & \Tsec  &  K  \\
 Tertiary effective temperature    &    $T_{\rm ter}$    & \Tter  &  K  \\

 
\noalign{\smallskip} \hline
\multicolumn{4}{c}{\emph{Derived fundamental parameters}} \\  
\hline \noalign{\smallskip}

 Primary mass      &    $M_{\rm pri}$    & \Mpri  &    M$_{\odot}$  \\
 Secondary mass    &    $M_{\rm sec}$    &  \Msec &    M$_{\odot}$  \\
 Primary radius    &    $R_{\rm pri}$    &  \Rpri  &    R$_{\odot}$  \\
 Secondary radius  &    $R_{\rm sec}$    &  \Rsec  &    R$_{\odot}$  \\
 Tertiary radius  &    $R_{\rm ter}$    &  \Rter  &    R$_{\odot}$  \\
 Primary luminosity    &    $L_{\rm pri}$    &  \Lpri &    L$_{\odot}$  \\
 Secondary luminosity    &    $L_{\rm sec}$    & \Lsec &    L$_{\odot}$  \\

 Primary surface gravity    &    $\log g_{\rm pri}$    & \loggpri &    (cm\,s$^{-2}$)  \\
 Secondary surface gravity    &    $\log g_{\rm sec}$    & \loggsec  &    (cm\,s$^{-2}$)  \\
 
 Mass sum     &    $M_{\rm pri} + M_{\rm sec}$    & \masssum &    M$_{\odot}$  \\
 Radius sum    &    $R_{\rm pri} + R_{\rm sec}$    & \radsum  &    R$_{\odot}$  \\
 
 
\noalign{\smallskip} \hline
\multicolumn{4}{c}{\emph{Derived radiative, orbital and rotational parameters}} \\  
\hline \noalign{\smallskip}

 Central surface brightness ratio in NGTS   &    $J_{\rm{NGTS}}$    &  \Jngts     &   \\
 Central surface brightness ratio in \TESS   &    $J_{\rm{\TESS}}$    &   \Jtess  &   \\
 Central surface brightness ratio in SPECULOOS $I+z'$ & $J_{\rm{SPECULOOS}}$    &     \Jspec   &   \\
 Central surface brightness ratio in SAAO \textit{V}  &    $J_{\rm{SAAO}}$    &     \Jsaao    &   \\
 Third light in NGTS   &    $L3_{\rm{NGTS}}$    &  \thirdngts     &   \\
 Third light in \TESS   &    $L3_{\rm{\TESS}}$    &   \thirdtess  &   \\
 Third light in SPECULOOS $I+z'$ & $L3_{\rm{SPECULOOS}}$    &     \thirdspec   &   \\
 Third light in SAAO \textit{V}  &    $L3_{\rm{SAAO}}$    &     \thirdsaao    &   \\
 
 Semi-major axis    &    $a$    &   \semimaj  &  R$_{\odot}$  \\
 Orbital inclination  &  $i$  & \orbinc  &  $^{\circ}$  \\
 Eccentricity    &    $e$    &  \eccen  &    \\
 Longitude of periastron    &    $\omega$    & \longperi  &    $^{\circ}$  \\

 Primary synchronised velocity      &    $V_{\rm pri ~ sync}$    &  \prisyncvel &  km\,s$^{-1}$  \\
 Secondary synchronised velocity    &    $V_{\rm sec ~ sync}$    & \secsyncvel  &  km\,s$^{-1}$  \\

 \noalign{\smallskip}  
 \hline  
 \end{tabular}
 \end{table*}}



\section{Results}
\label{sec:results}
\begin{figure}
	\includegraphics[width=\columnwidth]{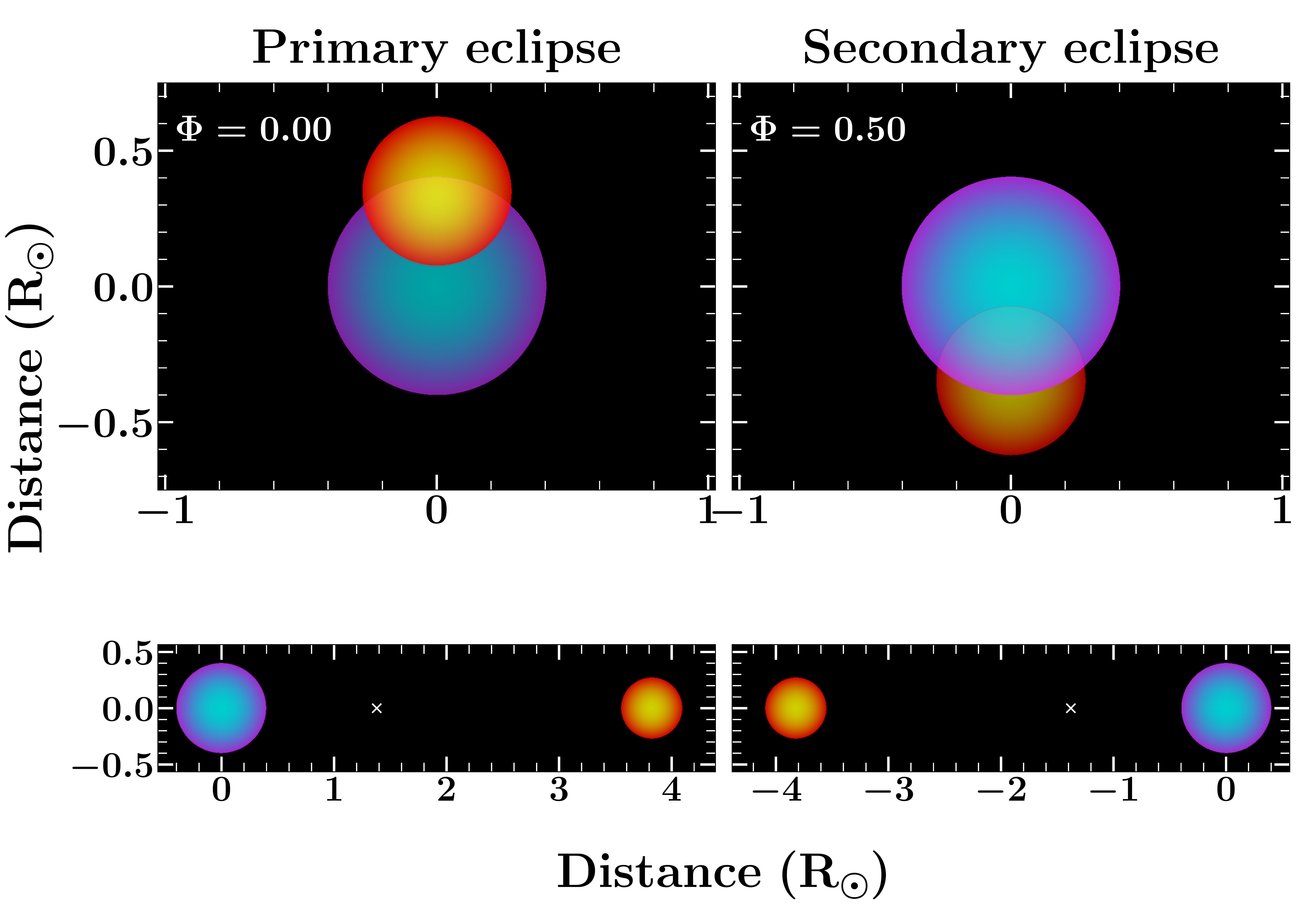}
    \caption{System geometry of \Nsys{} to scale at primary and secondary eclipse (left and right panels respectively). The primary star is shown in purple/blue and the secondary in orange/yellow, with the orbital phases labelled. The upper panels display our view with the primary star at the origin, whilst the lower panels give a side-on perspective in the orbital plane, perpendicular to a straight line joining the stars, to show the separation. The centre of mass is marked by a white cross.}
    \label{fig:geometry}
\end{figure}

We simultaneously modelled the NGTS, \TESS{}, SPECULOOS and SAAO light curves, UVES and HIRES RVs, and system SED with GP-EBOP. We used 400 `walkers' to explore parameter space in the MCMC. The chains were run for 400\,000 steps, the first 100\,000 steps were discarded as burn-in, and each chain was thinned based on the average auto-correlation time. \footnote{We also ran tests with the ensemble slice sampling MCMC method implemented in \zeus{} \citep{Karamanis2021}, with almost identical results.} 

Figures \ref{fig:LCs_GPEBOP} to \ref{fig:sed} display the model fits to the data. Figure \ref{fig:LCs_GPEBOP} shows each light curve with the global GP-EBOP model. The figure also shows, for NGTS and \tess{}, the light curves detrended with respect to the GP and phase-folded on the binary period, accompanied by close-ups of the eclipses. The RV orbit solution is shown in Figure \ref{fig:rvs}, phase folded on the orbital period of the binary. The sinusoidal curves indicate a negligible eccentricity, as expected given the period. The measured RVs of the tertiary component are also plotted. The derived systemic velocity is \velsys{} km s\textsuperscript{-1} (dashed gray line), and we measure the weighted-mean RV of the tertiary component as $5.63 \pm0.38$ km s\textsuperscript{-1}. These values are encouragingly similar to the estimate for the Blanco 1 cluster-centre RV ($5.78 \pm 0.10$ km s\textsuperscript{-1} \citetalias{GAIA2018}), and bode well for our assumption that the tertiary is physically associated with the binary. In Figure \ref{fig:sed} we plot the system SED. We show the BT-Settl model fit to the observed broadband magnitudes and the derived SEDs of each component. The system geometry at primary and secondary eclipse is depicted in \ref{fig:geometry}. We see a grazing eclipse, something in-keeping with the poor constraints on the radius ratio found prior to our SED modelling. The main parameters of the fit are given in the top section of Table \ref{tab:params}, with derived parameters in the middle and bottom sections.

We find the masses, radii and effective temperatures of the binary components in \Nsys{} to be: $M_{\rm pri}=\Mpri$ \msun, $M_{\rm sec}=\Msec$ \msun, $R_{\rm pri}=\Rpri$ \rsun, $R_{\rm sec}=\Rsec$ \rsun, $T_{\rm pri}=\Tpri$ K and $T_{\rm sec}=\Tsec$ K. For the tertiary, we find  $R_{\rm ter}=\Rter$ \rsun{} and $T_{\rm ter}=\Tter$ K. We note that our effective temperatures have a strong dependence on the particular stellar atmosphere models used (although our masses and radii do not). We present our main results using BT-Settl model atmospheres, but compare with the PHOENIX model atmospheres of \citet{Husser2013} in \S\ref{sub:BT_settle_Phoenix_comp}.


\section{DISCUSSION}
\label{sec:discussion}
\subsection{Mass--radius relation for low-mass EBs}
\begin{figure*}
	\includegraphics[width=\textwidth]{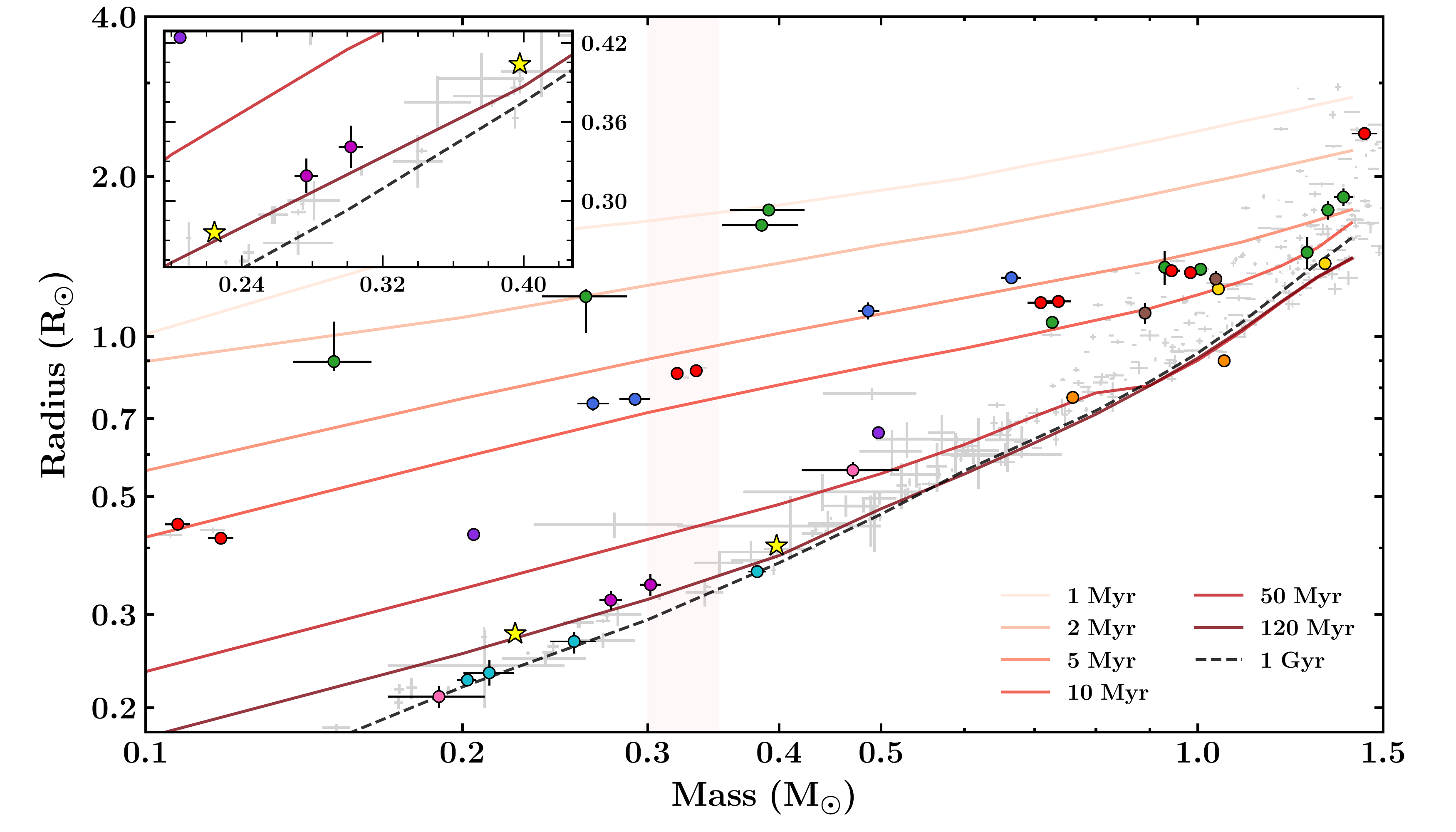}
    \caption{Mass--radius relation for detached, double-lined, stellar-mass EBs below 1.5 \msun. The coloured lines represent the solar metallicity isochrones of BHAC15 from 1 Myr to 1 Gyr (light-to-dark, top-to-bottom). Well-characterised EBs in sub-Gyr open clusters are coloured, while field EBs are shown in grey. \Nsys{} is plotted with yellow stars, with other cluster EBs shown in green (Orion), red (Upper Scorpius; including the eclipsing triple HD 144548), pink (NGC 1647), gold (Per OB2), magenta (Pleiades), orange (Hyades), cyan (Praesepe), brown (Upper Centaurus Lupus), blue (NGC 2264) and violet (32 Orionis Moving Group). The fully convective boundary is represented by a pink vertical bar. The cluster EBs (uncertainties $\lesssim10\%$; component masses < 1.5 \msun{}; ages < 1 Gyr) are as compiled by \citet{Gillen2017}, with subsequent additions from \citet{Chew2019}, \citet{David2019}, \citet{Murphy2020} and \citet{Gillen2020EB}.
    The field sample comprises the DEBCat catalogue (mass and radius uncertainties < 2\%; \citet{Southworth2015}); the close ($P < 10$ d) systems with M-dwarf primaries collected by \citet{Nefs2013} (non-detached systems removed); and additional EBs from \citet{Irwin2009, Irwin2018}, \citet{Stassun2014}, \citet{Zhou2015}, \citet{Dittmann2017}, \citet{Casewell2018} and \citet{Miller2021}. Inset: zoom on the region around \Nsys{}.}
    \label{fig:MR_EBs}
\end{figure*}

Figure \ref{fig:MR_EBs} shows the mass--radius relation for detached, double-lined, stellar-mass EBs below 1.5 \msun. The coloured lines represent the \citet{Baraffe2015} (hereafter BHAC15) isochrones from 1 Myr to 1 Gyr, and the data points show measurements for EBs in the field (grey) and in sub-Gyr open clusters (coloured; see figure caption for colour scheme). \Nsys{} is shown with yellow stars. The sub-Gyr cluster EBs represent some of the best tests of stellar evolution theory at low masses and young ages; \Nsys{} brings the total in this ensemble to 20. The inset to Figure \ref{fig:MR_EBs} shows a close-up of the region around \Nsys{}, and it is apparent that the primary has a larger radius than the three field stars most similar in mass. The inset also displays the only well-characterised low-mass EB from the Pleiades (HCG 76; \citealt{David2016}) (magenta markers), where the LDB age estimates of $125\pm{8}$ Myr \citep{Stauffer1998} and $112\pm{5}$ Myr \citep{Dahm2015} are similar to Blanco 1. Its components (like the \Nsys{} primary) prefer younger-than-canonical ages when compared with the BHAC15 isochrones.

\subsection{Context}
In addition to its main use as a test of stellar evolution theory, the properties of \Nsys{} place it within three interesting sub-groups of the known double-lined EB population: 1) those with low mass ratios 2) those with known tertiary companions and 3) those with components which span the fully convective boundary (see Figure \ref{fig:triples_conv_spanners}). Although membership of any one of these sub-groups is not an exceptionally rare trait, membership of all three is (to the best of our knowledge) unique for a well-characterised cluster EB. Binary mass ratios are relevant to the study of stellar evolution, with low-mass-ratio EBs ensuring that model predictions are tested over a wide range of masses for a single metallicity and age. The presence of tertiary companions is also relevant to stellar evolution and has been linked to particularly large model--observation discrepancies \citep{Stassun2014}, whilst differences in energy transport for components spanning the fully convective boundary should provide stringent tests of evolutionary models. We therefore discuss these topics, before comparing the measured and derived properties of \Nsys{} with stellar evolution models. 

\subsubsection{Mass ratios}
\label{sub:mass_ratios}
The distribution of binary mass ratios ($q=M_{\textrm{sec}}/M_{\textrm{pri}}$) should contain information about the components' formation and early evolution. If a protobinary forms within a collapsing molecular cloud core, the final masses will depend on how the stars accrete and interact with the surrounding material. One well-subscribed idea is that, in such a scenario, mass ratios will tend to be biased towards unity and that low-mass ratios will be rare for short-period systems, compared with longer-period binaries \citep{Bate1997, Bate2000, Young2015}. One of the reasons for this is that the specific angular momentum of infalling material is higher, relative to the binary, when the separation between the two stars is smaller, and accretion is preferentially directed towards the lower-mass secondary when that material has high angular momentum. At low angular momentum, gas falls towards the centre-of-mass of the system and so is accreted mainly by the primary, but with increasing angular momentum, circumstellar disks may form around primary and secondary, leading to more accretion by the secondary. With sufficient angular momentum, a circumbinary disk forms---with an inner edge closer to the secondary---and accretion will tend to drive the mass ratio towards one. The relative accretion rate of secondary to primary in the presence of a circumbinary disk is a strong function of the initial mass ratio, with low mass ratios heavily favouring the secondary \citep{Bate1997}. Alternatives to this mechanism have been proposed which instead favour accretion onto the primary \citep{Ochi2005, Hanawa2009, deValBorro2011}, but the assumed gas temperatures in those simulations could be too high to be representative of stellar binaries; at lower temperatures accretion would still favour the secondary \citep{Young2015}.

The other main reason for the expectation of few extreme mass ratios in close binary systems is one of dynamics. In the early stages of a binary system's life, interactions with other stars formed from the same or nearby cores are more likely than at later times, when the star-forming regions are dispersed. Such interactions, as demonstrated by simulations, are expected to lead to the ejection of the least massive component. This would naturally lead to an equalisation of masses and to massive stars being more likely to have close companions than lower-mass stars \citep{Bate2002}. Finally, higher mass-ratio binaries have higher binding energies and so are more resistant to disruption \citep{ElBadry2019}.

Observationally testing theories about binary mass ratios has been challenging historically. \citet{Duchene2013} note, in their review on stellar multiplicity, how the difficulties associated with the detection of low-mass companions has led to widely discrepant conclusions (see e.g. \citealt{Trimble1990}). Modern volume-limited surveys, with their large sample sizes, are promising means of achieving more reliability. In such studies (e.g. \citet{Raghavan2010, Moe2017, ElBadry2019}, significant excess twin fractions, $\mathcal{F}_{\rm{twin}}$, (\textit{twin} meaning $q>0.95$), at shorter periods are indeed found, in line with theoretical expectations.

In the low-mass domain, \citet{Bergfors2010} found evidence for a peak at $q\gtrsim0.7\textrm{--}0.8$ in the mass ratio distribution of mid-type M dwarfs (M 3.5--M 5.5), but not for early M dwarfs. \citet{Nefs2013} analysed the mass ratio distribution of known M-dwarf binaries, finding that over 80\% of stellar binaries have $q\ge 0.8$, although they noted how low-mass, low-luminosity companions may be unresolved in optical spectroscopy and so bias the distribution towards equal mass ratios. As \cite{ElBadry2019} explain, the bias against low-mass companions is a feature of all binary detection methods, e.g. low-mass stars induce weaker RV shifts for a given separation, contribute less light to observed spectra, create shallower eclipses, and are less likely to be detected as part of visual binaries. Additionally, the detection efficiency varies with primary mass and separation, which makes attempts to correct for incompleteness and bias all the more challenging. At even later spectral types (M 7--M 9.5), recent work, based on a large homogeneous sample from \citet{Ahmed2019}, has suggested that almost all unresolved binaries are likely to be twins \citep{Laithwaite2020}. However, the NGTS discovery of an M dwarf EB with mass ratio $q=0.14$ and $M_{\textrm{sec}}=0.08~\msun$ \citep{Acton2020b} highlights that extreme mass ratios do exist.

\begin{figure*}
	\includegraphics[width=\textwidth]{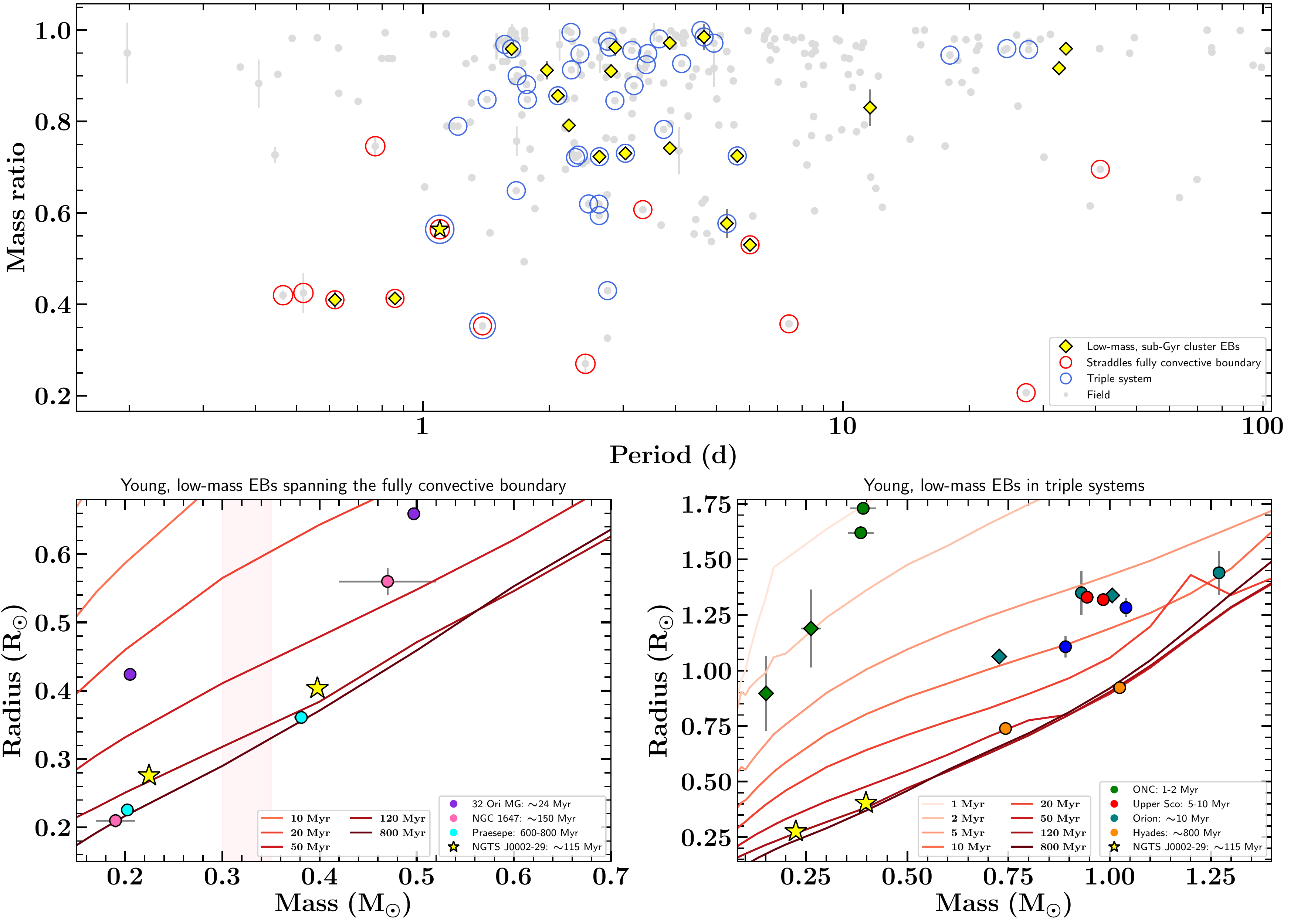}
    \caption{Top: Mass ratio vs orbital period for detached, double-lined, stellar-mass EBs. The well-characterised, low-mass, sub-Gyr-cluster systems collated by \citet{Gillen2017} and updated in \citet{Gillen2020EB} are plotted with yellow diamonds. The field EBs from \citet{Irwin2009, Irwin2018}, \citet{Stassun2014}, \citet{Nefs2013}, \citet{Zhou2015}, \citet{Dittmann2017}, \citet{Casewell2018}, \citet{Miller2021} and DEBCat \citep{Southworth2015} are plotted in grey. The red circles identify binaries where the components span the fully convective boundary, and the blue circles identify those with known tertiary companions. \Nsys{} is shown with a yellow star. Bottom left: The four well-characterised, double-lined cluster EBs whose components span the fully convective boundary (pink vertical bar), along with BHAC15 isochrones from 10--800 Myr. Bottom right: The well-characterised, double-lined cluster EBs with component masses < 1.5 \msun{}, which have known tertiary components, along with BHAC15 isochrones from 1--800 Myr.}
    \label{fig:triples_conv_spanners}
\end{figure*}

In the top plot of Figure \ref{fig:triples_conv_spanners}, we show binary mass ratio as a function of orbital period for detached, double-lined, stellar-mass EBs. There is visible clustering of systems towards $q\approx1.0$, with a median mass ratio of $q=0.92$. The mass ratio of \Nsys{} ($q=0.564\pm0.003$) is smaller than $\sim$95\% of the systems shown, and, whilst not extreme, does place it in a fairly sparsely-populated region of the diagram.

\subsubsection{Triple systems and the tertiary component}
\label{sub:tertiary}
\citet{Tokovinin2006} found that the vast majority (96\%) of solar-type spectroscopic binaries with periods shorter than 3 days have an additional companion, a result with recent corroboration from \citet{Laos2020}. In both studies, the frequency of triple systems was found to be a strong function of binary period, with tertiary companions absent for the majority of binaries with $P>6$ days. An obvious interpretation is that the tertiary companion plays a role in the creation of the closest binary systems. Current explanations of star formation preclude the in-situ formation of close binaries, because the latter stages of collapse proceed from a hydrostatic core of radius ${\sim}5$ au, which is resistant to further fragmentation \citep{Larson1969, Bate1998, Bate2011}. Fragmentation of collapsing regions of cold molecular clouds may lead to bound binary or multiple stellar systems if multiple collapse events occur within a turbulent parent core \citep{Ofner2010, Kratter2011}. Alternatively, gravitational instability within a protostellar disk may lead to fragmentation and the birth of additional companions \citep{Bonnell1994, Kratter2010}. In either of these scenarios of core or disk fragmentation, binary components separated by less than ${\sim}10$ au must have migrated inwards. 

Recent population synthesis work by \citet{Moe2018} concluded that the majority of close binaries with $P<10$ days form in the pre-main sequence (PMS), in agreement with observations, and derive from disk fragmentation followed by dynamical interactions of initially unstable triple systems, with significant energy dissipation in the disk, a mechanism consistent with the modelling of \citet{Tokovinin2020}. This is in contrast to orbital decay via Kozai-Lidov (KL) cycles and tidal friction in misaligned triples (e.g. \citealt{Eggleton2006, Fabrycky2007, Naoz2014}). 

Understanding the dynamic evolution of close binaries is of interest in studies of various astrophysical phenomena, e.g. binary mass exchange, mergers and type 1a supernovae, but also to the use of EBs as tests of stellar evolution theory. \citet{Stassun2014} showed how---for a sample of 13 benchmark PMS EBs---the stellar properties of systems with known tertiary companions were in much poorer agreement with the predictions of evolutionary models than those of lone binaries. They proposed that such discrepancies could be explained by the regular input of orbital energy from the tertiary to the binary, with tidal interactions between the binary components becoming significant if their separation was sufficiently reduced. The bottom-right plot in Figure \ref{fig:triples_conv_spanners} shows isochrones from BHAC15, and the well-characterised, sub-Gyr, low-mass EBs with known tertiary companions. Despite a possibly-inflated primary, \Nsys{} is one of the better-fit systems here. We also note that it is the shortest-period EB with a known tertiary companion (see upper panel).

At the age of Blanco 1, M dwarfs are not expected to have quite settled onto the main sequence, though their radii will not reduce much further. The chaotic interactions and migration of early PMS life which may affect systems in the \citet{Stassun2014} sample, where ages span approximately 1--20 Myr, would not necessarily be a feature of \Nsys{} at $\sim$115 Myr. Our modelling indicates that \Nsys{} is circularised ($e\approx0$). We also see OOE variability on (or very close to) the orbital period, suggesting synchronisation, which, from theory, is expected to occur before circularisation\footnote{Circularisation and synchronisation times are $t_{\textrm{circ}}\approx2$ Myr and $t_{\textrm{sync}}\approx0.03$ Myr by Eqs. 6.2 and 6.1 in \citealt{Zahn1977}, but we note the theory therein is based on stars with convective envelopes.}. If \Nsys{} is indeed circularised and synchronised, and if equatorial and orbital planes are aligned, then we expect tidal dissipation to be minimal with the binary in an equilibrium state \citep{Hut1981}, unless the tertiary's orbit is small or eccentric enough to interfere.

Using the MCMC samples for the radius and effective temperature of the tertiary companion from our global modelling, we derive a tertiary mass of  $M_{\rm ter}=\Mter~\msun$ from the BHAC15 stellar evolution models. This estimate was obtained by interpolating the models to compute a fine grid of the parameters (effective temperature and radius) at each mass, followed by a 2D cubic interpolation in log\,$T_\textrm{eff}$--\textit{R} space (using the \texttt{griddata} routine in \scipy{}) from our posterior distributions onto the grid, yielding a distribution of tertiary masses, from which we have quoted the median and 16th/84th percentiles. With this mass, the derived component luminosities, our measurements of the tertiary RVs, and the sensitivity of \gaia{}, we can attempt to put some loose constraints on the tertiary orbit. Given the derived luminosities of the tertiary and the binary, we would expect the tertiary to be resolved in \gaia{} for separations greater than $\sim$1 arcsec \citep{Brandeker2019}, which implies that the tertiary is within $\sim$240 au of the binary. Given the scatter and uncertainties of our tertiary RVs, and the timing of our observations spanning 60 days, we estimate that we would only be sensitive to the reflex orbit of the tertiary if its separation is less than $\sim$1--2 au from the binary, assuming a co-planar, circular orbit. That we do not see variations indicative of such a close orbit, leads to the conclusion that the tertiary is likely to orbit at a distance of $\sim$2--240 au. Additionally, we see no evidence for eclipse timing variations in our light curves; the eclipse minima are aligned with their predictions across the $\sim$1.5-year baseline of photometric observations. This is consistent with the most likely scenario of a hierarchical triple, where the tertiary is distant and low-mass.  


\subsubsection{The fully convective boundary}
\label{sub:convective_boundary}
The transition into the fully convective regime for stellar interiors is predicted to occur at around 0.3--0.35 \msun \citep{Dorman1989, Chabrier1997}. Fully convective main-sequence stars are considered to be the simplest stars to describe theoretically, being relatively insensitive to model input parameters \citep{Feiden2014b}, but stellar evolution models frequently struggle to match observations for these, as well as higher-mass, M dwarfs (e.g. \citet{Morales2009, Torres2010, Feiden2014b, Kesseli2018}). Magnetic activity is often favoured as a potential cause of the disagreement between models and predictions, due to inhibition of bulk convection or the creation of starspots, but it is by no means resolved \citep{Chabrier_2007, Feiden2014a, Feiden2014b, Morrell2019}. 

In the past, some work has highlighted a possible difference in the amount of deviation from stellar evolution models above and below the fully convective boundary, e.g. \citet{Morales2010} pointed to radii being much closer to theoretical models and less scattered for $M\lesssim0.35\,\msun$, with more scatter---but a larger deviation evident---for $M\gtrsim0.35\,\msun$.  Recently, others have found no difference above and below the boundary \citep{Parsons2018}, whilst \citet{Kesseli2018}, in their study of 88 rapidly rotating single M dwarf stars, found greater disparities between predicted and measured radii at the lowest masses (13\%--18\% for $0.08<M<0.18\,\msun$ compared with 6\% for $0.18<M<0.4\,\msun$), but also stated that there was no significant change in the amount of inflation compared to models \textit{across} the fully convective boundary. 

That there are different physics at play is less in doubt. When studying the effects of magnetic activity on low-mass stars, \citet{Chabrier_2007} showed that fully convective stars are quite insensitive to changes in the mixing length parameter (which, when reduced, leads to decreased convective efficiency in partially convective stars), but that they are significantly affected by spot coverage. Within this framework, the measured properties of the most-studied---but highly discrepant---fully convective EB, CM Dra, may be reconcilable with suitably-adjusted models \citep{Morales2010}. \citet{MacDonald2012} also fitted CM Dra to model predictions, but by invoking a magnetic inhibition parameter and suppression of convection, along with removing the effect of polar spots biasing radius values upwards in EB light curve modelling. In both cases there is, however, much uncertainty, e.g. whether the large coverage of polar spots and/or the required super-megagauss magnetic fields actually exist (see \citet{Feiden2014b} for a detailed discussion). 

It could be the case that the disagreement between models and observations has a different origin above and below the fully convective boundary due to the different physics involved, but there remains much to explain. Indeed, different physics does not necessarily manifest itself in all relations of interest. For example, despite the absence of a tachocline---the interface between radiative core and convective envelope, thought to be the location of magnetic field shearing and amplification in differentially rotating stars above the fully convective boundary (e.g. \citealt{Charbonneau2014})---lower-mass stars have been found to follow an activity--rotation relation which is indistinguishable from their partially convective counterparts \citep{Wright2016, Wright2018}. 

In addition to similarities in some of the observed effects of their magnetic dynamos, fully convective stars are seen to fit the smooth trends in mass--luminosity and radius--luminosity through M spectral types \citep{Demory2009}. As noted by \citet{Stassun2011}, it would appear that such stars are indifferent, in terms of energy generation and output, to changes in structure or energy transport within. However, in the mass--temperature and radius--temperature planes, the fully-convective transition zone lies in a region of substantial change, where both theory and observation---albeit with a $\sim$250 K offset (e.g. \citealt{Dupuy2010})---show there to be a large range of masses and radii for a small range of spectral types \citep{Chabrier2000, Stassun2011}. 

Narrowing the focus to open cluster EBs, there are, to our knowledge, only four well-characterised, double-lined, stellar-mass systems (\Nsys{} included) which span the fully convective boundary (see Figure \ref{fig:triples_conv_spanners}, bottom left). These systems ought to be especially stringent tests of stellar evolution models because, as well as having well-determined parameters and ages, different physics are relevant to each component, plus the low mass ratios are good tests of model isochrone gradients. Interestingly, in these cases we have two systems---those in Praesepe \citep{Gillen2017} and the 32 Orionis moving group \citep{Murphy2020}---where the masses and radii of both components agree well with (non-magnetic) model predictions, and another two systems---in Blanco 1 (this work) and NGC 1647 \citep{Hebb2006}---where it is the higher-mass component that appears to be inflated (although the secondary in NGC 647 is smaller than expected for its assumed cluster age of 150 Myr). Three out of these four systems (not Praesepe) also have very short periods ($\sim$1 day), meaning that they will be fast rotators (assuming spin-orbit synchronisation) and hence would be expected to exhibit enhanced magnetic activity. That their lowest-mass components do not appear to be inflated may be a clue that rotation-induced magnetic activity is not the explanation, or at least not the whole story, for those fully-convective stars which appear inflated compared with models. With observations of fully-convective stars which do and do not fit model radii predictions, the situation for stellar evolution modelling remains complex. It should be noted that models in all of the above four systems fail to predict the inferred temperatures of one or both components.


\subsection{Comparison with stellar evolution models}
We compare the fundamental parameters of \Nsys{} with the closest-to-cluster-metallicity predictions of the following stellar evolution models in the mass--radius and  $T_\textrm{eff}$\nobreakdash--log\,\textit{L} planes (MRD and HRD hereafter): BHAC15; MESA Isochrones and Stellar Tracks (MIST v1.2, with $v/v_{\textrm{crit}}=0.4$ \footnote{We have used the rotating set of MIST isochrones but note that these are equivalent to the non-rotating versions on the PMS.}; \citealt{Mist1, Mist2}); PAdova and TRieste Stellar Evolution Code  (PARSEC v.1.2S; \citealt{Parsec}); the standard and magnetic models of Feiden \citep{Feiden2016}; and Stellar Parameters of Tracks with Starspots (SPOTS; \citealt{Somers2015, SPOTS}). 

The BHAC15 models are an update to the models of \citet{Baraffe1998}, now using BT-Settl model atmospheres and updated surface boundary conditions. The MIST v1.2 models are based on the MESA (Modules for Experiments in Stellar Astrophysics) stellar evolution package. Version 1.2S of the PARSEC models updates the relation between the temperature and Rosseland mean optical depth ($T$\nobreakdash--$\tau$) for the outer boundary conditions to that from the BT-Settl model atmospheres. Also included in v1.2S is a shift in the $T$\nobreakdash--$\tau$ relations to reproduce the observed mass–radius radius relation of low-mass dwarf stars \citep{Chen2014}. We note that this shift means that the v1.2S models are not a direct test of the underlying stellar evolution theory. \footnote{We include the PARSEC v1.2 models as they are commonly used in the literature and give quite different predictions to other models in the region of parameter space relevant to this work.} The Feiden models are based on the Dartmouth Stellar Evolution Program (DSEP; \citealt{Dartmouth}), and were further developed in \citet{Feiden_models2012, Feiden2013} and \citet{Feiden2016} to include the effect of magnetic fields. Magnetic fields act to inhibit convection and hence slow PMS contraction, which generally results in older age predictions compared to non-magnetic models. The SPOTS models use the Yale Rotating Evolution Code (YREC) and incorporate the structural effects of starspots. The effects of spots in these models are to suppress the rate of convective energy transport in the stellar interior and to alter the average pressure and temperature at the model photosphere. The SPOTS models are divided into two zones: spotted and un-spotted, each with an associated temperature at a given layer. The model temperature at any given radius within the star is then determined by summing the fluxes of the spotted and un-spotted regions. Model isochrones are available for six different spot surface covering fractions; our comparisons use $f=0\%$ and $f=17\%$.

\begin{figure*}
  \begin{minipage}[c]{0.75\textwidth}
    \includegraphics[width=\textwidth]{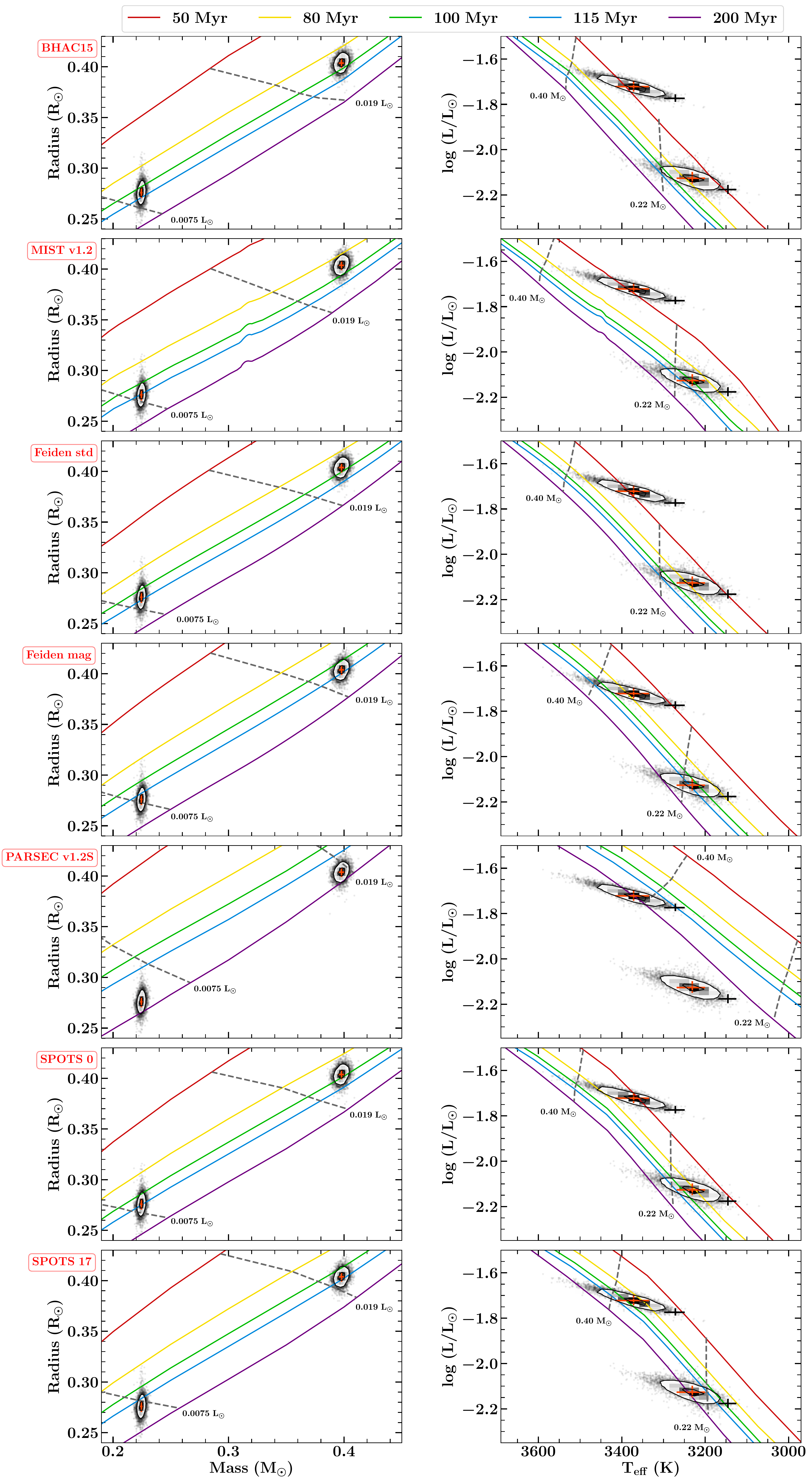}
  \end{minipage}\hfill
  \begin{minipage}[c]{0.25\textwidth}
    \caption{
       Comparison of the fundamental properties of \Nsys{} to the predictions of the BHAC15, MIST v.1.2, Feiden std, Feiden mag, PARSEC v1.2S, SPOTS 0 and SPOTS 17 stellar evolution models in the the mass--radius and  $T_\textrm{eff}$--log\,\textit{L} planes. Isochrones at 50 (red), 80 (yellow), 100 (green), 115 (blue) and 200 (purple) Myr are plotted, along with isolumes (mass--radius plane) and evolutionary tracks ($T_\textrm{eff}$--log\,\textit{L} plane) as grey dashed lines. The median and 1-sigma error bars from our posterior distributions are shown by orange crosses. The full distributions (grey dots), and the 1- and 2-sigma contours of the 2D distributions (39\% and 86\% of the volume) are plotted using the contour function in \corner{} \citep{corner}. Note that for the asymmetric distributions (in this case $T_\textrm{eff}$ and log\,\textit{L}), the 1D error bars do not align with the 2D 1-sigma contours. We also illustrate the effect of using PHOENIX model atmospheres by plotting black crosses in the $T_\textrm{eff}$--log\,\textit{L} plane. The masses and radii derived using the PHOENIX models are almost identical to the main results and so are not shown in the mass--radius plane.} \label{fig:all_models}
  \end{minipage}
\end{figure*}

In Figure \ref{fig:all_models}, we compare the properties of \Nsys{} to the stellar evolution models described above. In each panel, the coloured lines represent five isochrones from 50--200 Myr, and the location of \Nsys{} is shown with orange crosses. The grey dashed lines are evolutionary tracks at constant luminosity (in the MRD) and constant mass (in the HRD) for the values we have derived. In the HRD, we also plot (with black crosses) the location of \Nsys{} as determined when using the PHOENIX (see \S\ref{sub:BT_settle_Phoenix_comp}), rather than BT-Settl model atmospheres. Table \ref{tab:ages} compares model-predicted ages from the MRD and HRD. These estimates were arrived at by an equivalent procedure to that described in \S\ref{sub:tertiary} for the derivation of the tertiary mass, but in this case we used model isochrones rather than evolutionary tracks. We interpolated the models to compute a fine grid of masses, radii, effective temperatures and luminosities at each age, using isochrones at 20, 30, 40, 50, 80, 100, 120 and 200 Myr---a sampling density based on the finest available to all models.  This was followed by a 2D cubic interpolation \footnote{We find that the 2D interpolation works best in log\,$T_\textrm{eff}$--log\,\textit{L} space for the HRD.} from our posterior distributions onto the grid, to yield a distribution of ages. We do not give HRD age estimates based on the PARSEC models, because nearly all the data points fall well beyond the zero-age main sequence.

We find that the binary components appear coeval (within the 1-sigma error bars) in the MRD for the magnetic Feiden and PARSEC models, but component ages do not agree for any model in the HRD. There is greater uncertainty in the ages derived from the HRD than the MRD, which is a consequence of the measured masses and radii being better constrained than the effective temperatures and luminosities. It is evident from Table \ref{tab:ages} that, with the exception of the secondary in the SPOTS 17 models, age predictions based on the HRD are systematically younger than those based on the MRD, with the discrepancy being greater for the primary. The primary ages are also younger than the secondary ages in both the MRD and HRD. This could be interpreted as the primary, rather than the fully-convective secondary, being subject to inflation. The magnetic Feiden and SPOTS 17 models, as expected, produce older age estimates than the other models (PARSEC excluded), all of which are non-magnetic. They are also the only models whose MRD age predictions are both consistent with the LDB age of $115\pm 10$ Myr. In the HRD, the predicted age of the secondary for the magnetic Feiden models and the primary for the SPOTS 17 models are consistent with the LDB age. The primary and secondary ages between MRD and HRD are both consistent within the uncertainties for SPOTS 17, whereas it is only the secondary ages which are consistent for the magnetic Feiden models. One can see, when comparing the magnetic Feiden models and SPOTS 17 models in Figure \ref{fig:all_models}, that the posterior distributions lie in almost identical positions in the MRD, whilst appearing older for SPOTS 17 in the HRD. It is interesting to see how current magnetic stellar evolution models differ in their predictions, whilst bearing in mind that the Feiden models focus on how magnetic activity affects bulk convection, as opposed to the impact of starspots. We note that we do not account for spots explicitly in the GP-EBOP model.

We can also look at the plotted evolutionary tracks and observe whether our derived luminosities (in the MRD) and masses (in the  HRD) are consistent with model predictions. In the MRD, no models give a good match to both components, although SPOTS 17 comes closest. The isolumes, in most cases, intersect the distributions for the secondary between the 1- and 2-sigma regions, but not for the primary. In the HRD, the evolutionary tracks for the secondary are, again, reasonably well-matched to our observations, with the magnetic models (Feiden mag in particular) doing the best. However, all models (except PARSEC) underpredict the primary mass. This highlights how the estimation of stellar masses for young, low-mass objects from model isochrones and observational HR diagrams can be problematic. Overall, we find that the predictions of the magnetic models (Feiden mag and SPOTS 17) are a better match to our measurements than those of the non-magnetic models.

\subsection{Activity}
There are X-ray observations of \Nsys{} in the literature by \textit{ROSAT} \citep{Micela1999} and \textit{XMM-Newton} \citep{Pillitteri2003, Pillitteri2004, Pillitteri2005}, in which the system is referred to either by its 2MASS ID, or the designation BLX 24. The X-ray properties of the system, as given in \citet{Pillitteri2004}, are $\textrm{log~flux} = -13.68~\textrm{erg s}^{-1}\textrm{cm}^{-2}$; $\textrm{log}\,L_{X} = 29.20~\textrm{erg s}^{-1}$; and $\textrm{log} \frac{L_{X}}{L_{\rm{bol}}} = -2.31$. It is found to be an X-ray variable in \citet{Pillitteri2005}, although they note that its proximity to a chip gap and to another faint source might have influenced the evaluation of variability and the light curve.

In our HIRES spectrum, we see H$\alpha$ and H$\beta$ emission, and filled-in absorption features of the Ca{\footnotesize II} Infrared Triplet (8498/8542/8662 \AA), indicative of magnetic activity. We do not see any evidence of lithium at 6708 \AA{} in the HIRES or UVES spectra, but it is not expected for mid M dwarfs in Blanco 1 \citep{Juarez2014}. We fit the H$\alpha$ emission profiles using the same method as applied to the broadening functions in \S\ref{sub:rvs}, but with three Voigt functions in place of Gaussians. We measure equivalent widths of EW\textsubscript{pri}(H$\alpha\approx-4$ \AA) and EW\textsubscript{sec}(H$\alpha\approx-1$ \AA), but note that the three peaks are close in velocity space and so blending is an issue. We estimate measurement uncertainties of 5--10\% for the EWs, based on the values obtained in MCMC runs with different constraints. 

In order to compare H$\alpha$ emission between stars of different intrinsic luminosities, it is common to use the $L_{\rm{H\alpha}}/L_{\rm{bol}}$ metric (e.g. \citet{Walkowicz2004,Douglas2014,Newton2017}). This is defined as $L_{\rm{H\alpha}}/L_{\rm{bol}}=\rm{EW}_{\rm{H\alpha}}\times \it{f}_{\rm{0}}/f_{\rm{bol}}$, where ${f}_{\rm{0}}$ is the continuum flux for the line. ${f}_{\rm{0}}/f_{\rm{bol}}$ ($\chi$ hereafter) can either be calculated from high quality data and bolometric corrections, or from model atmospheres \citep{Reiners2007,Stassun2012,Douglas2014}. Using BT-Settl model atmospheres and $T_{\rm{eff}}$, log \textit{g} and metallicity values closest to the binary components, and taking the continuum flux to be the mean flux between 6550--6560 \AA{} and 6570--6580 \AA, we calculate log$\frac{L_{\rm{H\alpha}}}{L_{\rm{bol}}}=-3.81\pm{0.03}~\rm{and}~-4.87\pm{0.03}$ for primary and secondary respectively. \citet{Douglas2014} provide empirical spectral-type--$\chi$ relations and PHOENIX model $T_{\rm{eff}}$--$\chi$ relations, from which we find similar (though marginally larger) values. From the H$\alpha$ indicator, it appears that the primary is more active than the secondary, which could explain its apparently greater inflation. This is, however, based upon a single-epoch, with blended emission peaks, so additional spectra (ideally closer to quadrature) would be desirable.

\citet{Stassun2012} give empirical relations for predicting the amount by which the effective temperatures and radii of low-mass stars are changed due to chromospheric activity. They base the relations on a large set of low-mass field stars with H$\alpha$ measurements and a smaller set of low-mass EBs with X-ray activity measurements, from which they infer H$\alpha$ activity. Using our calculated $L_{\rm{H\alpha}}/L_{\rm{bol}}$ values, the relations give $\Delta T_{\rm{eff,pri}}=-5\pm{1}\%$, $\Delta T_{\rm{eff,sec}}=-2\pm{1}\%$, $\Delta R_{\rm{pri}}=10\pm{1}\%$ and $\Delta R_{\rm{sec}}=-2\pm{2}\%$\footnote{Radius deflation factors are probably unphysical. \citet{Stassun2012} say that offsets should simply approach zero at very low activity levels.}. For the non-magnetic models, the primary does not appear inflated by as much as 10\% in the MRD, unless the system age is $\sim$200 Myr. Rather, the average inflation factor we observe at the nominal system age of $\sim$115 Myr is $\sim$4\%. However, we do see an un-inflated secondary, in agreement with the empirical relations. Finally, shifting the primary by $\Delta \rm{T_{eff}}\sim 170~K$ ($\sim$5\%) and the secondary by $\Delta \rm{T_{eff}}\sim 65~K$ ($\sim$2\%) in the HRD would bring them into reasonable agreement with most non-magnetic models at $\sim$115 Myr. It should be noted that there is significant scatter in the \citet{Stassun2012} relations. 

\subsection{Age of \Nsys{}}
As stated in \S\ref{sec:intro}, a number of age estimates have been made for Blanco 1 over the past $\sim$25 years: $90\pm 25$ Myr based on H$\alpha$ emission \citep{Panagi1997}; LDB ages of $132\pm 24$ Myr and $115\pm 10$ Myr \citep{Cargile2010, Juarez2014}; $146\pm 14$ Myr based on gyrochronology \citep{Cargile2014}; and ${\sim}100$ Myr based on isochrone fitting \citepalias{Zhang2020}. \citetalias{Zhang2020} point out that the LDB age, adopted in \citetalias{GAIA2018} following a good fit to the lower main sequence, ought to be revisited. In the age of \gaia{}, there is good reason for such a study, because, out of the fourteen stars taken to be Blanco 1 members in \citet{Juarez2014}, only three (all bright objects) appear in the \citetalias{GAIA2018} and \citetalias{Zhang2020} Blanco 1 members lists. The majority of stars in the LDB study are of course faint, making confirmation of cluster membership more difficult, and is hence a possible reason for them being filtered out in \citetalias{GAIA2018} and \citetalias{Zhang2020}. Another reason could be that they are binaries and have high astrometric jitter. Table \ref{tab:ages} shows the isochronal ages we derive for \Nsys{} from the MRD and HRD. With the exception of the higher-than-expected PARSEC ages, we find MRD ages of $\sim$90--115 Myr from non-magnetic models and $\sim$110--125 Myr from magnetic models. HRD ages typically appear younger by $\sim$15--50 Myr.

\subsection{Distance to \Nsys{}}
\citetalias{GAIA2018} determined the parallax of the Blanco 1 cluster centre to be $4.216\pm0.003$ mas, equivalent to a distance of $237.19\pm0.17$ pc. \citet{Bailer2018} caution against directly converting parallaxes of individual objects into distances for those stars---a large majority in \gaia{}---where the relative uncertainty on the parallax is greater than 10--20\%. The formal relative uncertainty on the \Nsys{} parallax in both DR2 and eDR3 is $\sim$7\%, but, this is very large compared to stars of similar brightness and on-sky position, and is likely underestimated \citep{ElBadry2021}. There is also a large difference between the catalogue parallaxes, $3.9730\pm0.2615$ mas ($252\pm17$ pc; DR2) and $3.7976\pm0.2555$ mas ($263\pm18$ pc; eDR3) \footnote{Catalogue parallaxes are also subject to a small zero point offset. The provisional correction function in \citet{Lindegren2020L} suggests the eDR3 parallax of \Nsys{} is too small by $\sim$0.043 mas}, and the \citetalias{GAIA2018} parallax of $4.616\pm0.149$ mas ($217\pm7$ pc). The  \citetalias{GAIA2018} parallax accounts for the measured proper motion, the parallax and space motion of the cluster centre, and the position of the star on the sky relative to the projection of the cluster centre (see also \citet{gaia2017}). Such `kinematically improved' parallaxes are refinements for the vast majority of cluster members, but the fact that \Nsys{} is a multi-star system is problematic. Indeed, the \gaia{} astrometry is undoubtedly perturbed, as is evidenced by large error bars on the astrometric parameters and a large re-normalised unit weight error (RUWE = 4.15), the recommended statistical indicator for the reliability and quality of \gaia{} astrometry \citet{RUWE}. In the light of the work of \citet{Belokurov2020} and \citet{Stassun2021}, it is clear that the tertiary is the most-likely cause of the astrometric perturbations. 

We determine a distance to \Nsys{} of $228\pm6$ pc. This result comes out of our global modelling, where the SED model fluxes are scaled by the solid angle subtended by the stars at the fitted distance. We placed a Jeffreys prior on the parallax over a range of 3--6 mas, initialised at the \citetalias{GAIA2018} value for the cluster centre.

\subsection{Differences between using BT-Settl and PHOENIX atmosphere models}
\label{sub:BT_settle_Phoenix_comp}
In order to test the effect of the stellar atmosphere model used, we also modelled the system using PHOENIX model atmospheres. The results from this run yielded almost identical masses and radii (and uncertainties) to those obtained using the BT-Settl model atmospheres, but with effective temperatures which were ${\sim}90$ K cooler, differing by ${\sim}1.5\sigma$. The corresponding distance derived was smaller by ${\sim}15$ pc. Similar differences between these PHOENIX and BT-Settl model atmospheres have also been found in other studies of young, low-mass EBs, e.g. \citet{Murphy2020} found that the PHOENIX models gave effective temperatures which were ${\sim}30$ K cooler in their study of a ${\sim}24$ Myr-old system, whilst \citet{Gillen2020EB} found temperatures to be cooler by ${\sim}160$ K and ${\sim}125$ K for the two <10 Myr-old systems analysed therein. The effect of the lower temperatures and luminosities in the context of stellar evolution models is illustrated in Figure \ref{fig:all_models}, where the black crosses in the HRDs are shifted down and to the right, relative to the main results. This shift corresponds to age predictions being between 11 and 85 Myr younger for the primary and between 23 and 84 Myr younger for the secondary.

{\renewcommand{\arraystretch}{1.5}
\begin{table}
	\centering
	\caption{Isochronal ages of \Nsys{} in the mass--radius (MR) and Hertzsprung--Russell (HR) diagrams.}
	\label{tab:ages}
	\begin{tabular}{lcccc}
		\hline \hline
Model &\multicolumn{2}{c}{MRD age (Myr)} & \multicolumn{2}{c}{HRD age (Myr)} \\
    \noalign{\smallskip} 
   & Primary &	Secondary   & Primary & Secondary\\
		\hline
BHAC15     & $95\pm4$ &	$106\,^{+8}_{-6}$   & $44\,^{+10}_{-7}$ & $71\,^{+17}_{-12}$\\
MIST v1.2     & $90\pm5$ &	$114\,^{+10}_{-7}$   & $47\,^{+6}_{-5}$ & $100\,^{+16}_{-13}$\\
Feiden (std) & $96\,^{+5}_{-4}$ &	$110\,^{+9}_{-6}$   & $46\,^{+10}_{-8}$ & $74\,^{+16}_{-12}$\\
Feiden (mag) & $109\,^{+7}_{-5}$ & $124\,^{+12}_{-8}$   & $73\,^{+17}_{-11}$ & $110\,^{+31}_{-15}$\\
PARSEC v1.2S & $166\,^{+17}_{-15}$ &	$176\,^{+12}_{-11}$   &  & \\
SPOTS 0 & $97\pm4$ &	$112\,^{+9}_{-7}$   & $54\,^{+12}_{-9}$ & $85\,^{+20}_{-15}$\\
SPOTS 17 & $109\,^{+7}_{-5}$ &	$127\,^{+13}_{-9}$   &$90\,^{+15}_{-14}$ &$179\,^{+57}_{-50}$\\
		\hline
	\end{tabular}
\end{table}}

\subsection{Another EB in Blanco 1?}
\label{sub:HD224113}
\Nsys{} is the first well-characterised, low-mass EB in Blanco 1, but there exists a high-mass system which could also be a cluster member. The bright, early-type (B7V + B9V), double-lined EB, HD 224113 (\gaia{} eDR3 ID: 2314213698611350144), was characterised by \citet{Haefner1987}. Its \gaia{} parallax is consistent with Blanco 1, and despite a large range of recorded centre-of-mass velocities (ranging between $-$2.6 and +10.4 km s\textsuperscript{-1}), these would not rule out cluster membership. It is not listed as a member in \citetalias{GAIA2018} or \citetalias{Zhang2020}, but perturbed astrometry induced by its binary nature is a plausible reason for its absence. Its on-sky position would place it as a moderate outlier amongst the \citetalias{GAIA2018} cluster members, although not an outlier within the proposed \citetalias{Zhang2020} list. However, its proper motion, as measured by \gaia{} DR2/eDR3, would make it a more extreme outlier, compared with the \gaia{} DR2-confirmed members. Thus, to the best of our knowledge, the current census of EBs in Blanco 1 consists of either one or two systems, with \Nsys{} potentially the only well-characterised EB in the cluster.

\section{CONCLUSIONS}
\label{sec:conclusions}
We have presented the identification and characterisation of \Nsys{} as an EB in the $\sim$115 Myr old Blanco 1 open cluster. The star system is an M-dwarf triple, consisting of a detached, double-lined EB, whose components span the fully convective boundary, and a low-mass tertiary companion.

We simultaneously modelled light curves, RVs and the system SED with GP-EBOP to yield high-precision parameter estimates, including masses to <1\% and radii to <2\%. We applied light ratio constraints from our UVES spectra, propagated through the SED model into the light curve bands and hence into the eclipse modelling, in order to break the degeneracy between radius ratio, inclination and surface brightness ratio. The dataset was composed of our NGTS discovery light curve, \tess\ observations, follow-up photometry from SPECULOOS and SAAO, and spectra from VLT/UVES and Keck/HIRES.

We found that the binary components travel on circular orbits around their common centre of mass in $P_{\rm orb} =\period$ days, and have masses  $M_{\rm pri}=\Mpri$ M$_{\odot}$ and $M_{\rm sec}=\Msec$ M$_{\odot}$, radii $R_{\rm pri}=\Rpri$ R$_{\odot}$ and $R_{\rm sec}=\Rsec$ R$_{\odot}$, and effective temperatures $T_{\rm pri}=\Tpri$ K and $T_{\rm sec}=\Tsec$ K. We compared these properties to the predictions of seven stellar evolution models, revealing a possibly-inflated primary. We found MRD ages of $\sim$90--115 Myr from non-magnetic models and $\sim$110--125 Myr from magnetic models.

\Nsys{} is currently the only well-characterised EB of known age which has both a confirmed tertiary companion and components which straddle the fully convective boundary. Furthermore, it is one of only two well-characterised, low-mass EBs with an age close to $\sim$115 Myr, which makes the system a benchmark addition to the growing list of low-mass, sub-Gyr EBs that constitute some of the strongest observational tests of present and future stellar evolution theory at low masses and young ages.

\section*{ACKNOWLEDGEMENTS}
This research is based on data collected under the NGTS project at the ESO La Silla Paranal Observatory. The NGTS facility is funded by a consortium of institutes consisting of 
the University of Warwick,
the University of Leicester,
Queen's University Belfast,
the University of Geneva,
the Deutsches Zentrum f\" ur Luft- und Raumfahrt e.V. (DLR; under the `Gro\ss investition GI-NGTS'),
the University of Cambridge, together with the UK Science and Technology Facilities Council (STFC; project reference ST/M001962/1). This work is also based on observations collected at the European Southern Observatory under ESO programme 0103.C-0902(A) and at the W. M. Keck Observatory, which is operated as a scientific partnership among the California Institute of Technology, the University of California, and the National Aeronautics and Space Administration.
This research received funding from the European Research Council (ERC) under the European Union's Horizon 2020 research and innovation programme (grant agreement n$^\circ$ 803193/BEBOP), and from the Science and Technology Facilities Council (STFC; grant n${^\circ}$ ST/S00193X/1 and ST/S00305/1).

GDS gratefully acknowledges support by an STFC-funded PhD studentship and thanks Simon Hodgkin and Floor van Leeuwen for helpful discussion concerning the \gaia{} astrometry.
EG gratefully acknowledges support from the David and Claudia Harding Foundation in the form of a Winton Exoplanet Fellowship.
MNG acknowledges support from MIT's Kavli Institute as a Juan Carlos Torres Fellow.
LD is an F.R.S.-FNRS Postdoctoral Researcher.
Finally, we would like to thank the referee, John Southworth, for his helpful and very positive report.

\section*{Data Availability}
The data underlying this article will be shared on reasonable request to the corresponding author.



\bibliographystyle{mnras}
\bibliography{paper} 








\bsp	
\label{lastpage}
\end{document}

%% file: commands.tex
\newcommand{\tess}{{\it TESS}}

\newcommand{\gaia}{{\it Gaia}}

\newcommand{\NGTS}{NGTS}
\newcommand{\TESS}{{\it TESS}}

\newcommand{\masy}{mas\,y$^{-1}$}

\newcommand{\msun}{\mbox{$\mathrm{M}_{\odot}$}}
\newcommand{\rsun}{\mbox{$\mathrm{R}_{\odot}$}}

\newcommand{\emcee}{\texttt{emcee}}
\newcommand{\celerite}{\texttt{celerite}}
\newcommand{\celeritetwo}{\texttt{celerite2}}
\newcommand{\george}{\texttt{george}}
\newcommand{\eleanor}{\texttt{eleanor}}
\newcommand{\Python}{\texttt{Python}}
\newcommand{\astropy}{\texttt{astropy}}
\newcommand{\PyAstronomy}{\texttt{PyAstronomy}}
\newcommand{\scipy}{\texttt{scipy}}
\newcommand{\corner}{\texttt{corner}}
\newcommand{\zeus}{\texttt{zeus}}

%% file: properties.tex
\newcommand{\NRA}{\mbox{$00^{\rmn{h}} 02^{\rmn{m}} 48\fs4549$}} 
\newcommand{\NDec}{\mbox{$-29\degr 53\arcmin 53\farcs 8898$}} 
\newcommand{\NpropRA}{\mbox{$18.978\pm0.231$}} 
\newcommand{\NpropDec}{\mbox{$3.615\pm0.203$}} 
\newcommand{\Nparallax}{\mbox{$3.7976\pm0.2555$}} 

\newcommand{\PANgmag}{$18.0980 \pm 0.0077$}
\newcommand{\PANrmag}{$16.9314 \pm 0.0025$}
\newcommand{\PANimag}{$15.5682 \pm 0.0029$}
\newcommand{\PANzmag}{$14.9537 \pm 0.0018$}
\newcommand{\PANymag}{$14.6535 \pm 0.0027$}
\newcommand{\NGAIAGmag}{$16.40242 \pm 0.00453$}
\newcommand{\NGAIABmag}{$17.78037 \pm 0.01058$}
\newcommand{\NGAIARmag}{$15.08023 \pm 0.00467$}
\newcommand{\NJmag}{$13.434 \pm 0.029$}
\newcommand{\NHmag}{$12.829 \pm 0.023$}
\newcommand{\NKmag}{$12.556 \pm 0.024$}
\newcommand{\NWmag}{$12.444 \pm 0.023$}
\newcommand{\NWWmag}{$12.271 \pm 0.023$}
\newcommand{\APgmag}{$18.331 \pm 0.254$}
\newcommand{\APrmag}{$16.903 \pm 0.211$}
\newcommand{\APimag}{$15.509 \pm 0.222$}
\newcommand{\SKYgmag}{$17.9057 \pm 0.0440$}
\newcommand{\SKYrmag}{$16.9722 \pm 0.0124$}
\newcommand{\SKYimag}{$15.4421 \pm 0.0048$}
\newcommand{\SKYzmag}{$14.8559 \pm 0.0084$}

\newcommand{\PANgfl}{$(2.669 \pm 0.019)\times10^{-16}$}
\newcommand{\PANrfl}{$(4.779 \pm 0.011)\times10^{-16}$}
\newcommand{\PANifl}{$(1.136 \pm 0.003)\times10^{-15}$}
\newcommand{\PANzfl}{$(1.510 \pm 0.003)\times10^{-15}$}
\newcommand{\PANyfl}{$(1.616 \pm 0.004)\times10^{-15}$}
\newcommand{\NGAIAGfl}{$(7.8466 \pm 0.0328)\times10^{-16}$}
\newcommand{\NGAIABfl}{$(3.2435 \pm 0.0316)\times10^{-16}$}
\newcommand{\NGAIARfl}{$(1.2252 \pm 0.0053)\times10^{-15}$}
\newcommand{\NJfl}{$(1.324 \pm 0.035)\times10^{-15}$}
\newcommand{\NHfl}{$(8.371 \pm 0.177)\times10^{-16}$}
\newcommand{\NKfl}{$(4.067 \pm 0.090)\times10^{-16}$}
\newcommand{\NWfl}{$(8.693 \pm 0.184)\times10^{-17}$}
\newcommand{\NWWfl}{$(3.002 \pm 0.064)\times10^{-17}$}
\newcommand{\APgfl}{$(2.322 \pm 0.543)\times10^{-16}$}
\newcommand{\APrfl}{$(4.977 \pm 0.967)\times10^{-16}$}
\newcommand{\APifl}{$(1.220 \pm 0.250)\times10^{-15}$}
\newcommand{\SKYgfl}{$(2.964 \pm 0.120)\times10^{-16}$}
\newcommand{\SKYrfl}{$(4.756 \pm 0.054)\times10^{-16}$}
\newcommand{\SKYifl}{$(1.207 \pm 0.005)\times10^{-15}$}
\newcommand{\SKYzfl}{$(1.496 \pm 0.012)\times10^{-15}$}

\newcommand{\Nsys}{NGTS J0002-29}
\newcommand{\period}{\mbox{$1.09800524 \pm 0.00000038$}}%
\newcommand{\epoch}{\mbox{$2457998.65772 \pm 0.00015$}}%
\newcommand{\rsum}{\mbox{$0.1777\pm 0.0017$}}%
\newcommand{\rratio}{\mbox{$0.684\pm 0.015$}}%
\newcommand{\cosi}{\mbox{$0.0913\,^{+0.0028}_{-0.0033}$}}%
\newcommand{\recosw}{\mbox{$0.0071\,^{+0.0062}_{-0.0050}$}}%
\newcommand{\resinw}{\mbox{$-0.005\pm 0.031$}}%
\newcommand{\velsys}{\mbox{$5.39\pm 0.18$}}%
\newcommand{\rvpri}{\mbox{$63.31\pm 0.26$}}%
\newcommand{\rvsec}{\mbox{$112.17\pm 0.38$}}%
\newcommand{\Dist}{\mbox{$227.8\,^{+5.8}_{-5.5}$}}%
\newcommand{\Av}{\mbox{$0.045\,^{+0.086}_{-0.035}$}}%
\newcommand{\Tpri}{\mbox{$3372\,^{+44}_{-37}$}}%
\newcommand{\Tsec}{\mbox{$3231\,^{+38}_{-31}$}}%
\newcommand{\Tter}{\mbox{$3183\,^{+93}_{-104}$}}%
\newcommand{\Mpri}{\mbox{$0.3978\pm 0.0033$}}%
\newcommand{\Msec}{\mbox{$0.2245\pm 0.0018$}}%

\newcommand{\Mter}{\mbox{$0.16\pm 0.03$}}%

\newcommand{\Rpri}{\mbox{$0.4037\pm 0.0048$}}%
\newcommand{\Rsec}{\mbox{$0.2759\pm 0.0055$}}%
\newcommand{\Rter}{\mbox{$0.25\pm 0.03$}}%
\newcommand{\Lpri}{\mbox{$0.019\pm 0.001$}}%
\newcommand{\Lsec}{\mbox{$0.0075\pm 0.0005$}}%
\newcommand{\loggpri}{\mbox{$4.826\pm 0.010$}}%
\newcommand{\loggsec}{\mbox{$4.908\pm 0.017$}}%
\newcommand{\masssum}{\mbox{$0.6223\pm 0.0048$}}%
\newcommand{\radsum}{\mbox{$0.6797\,^{+0.0064}_{-0.0069}$}}%
\newcommand{\Jngts}{\mbox{$0.744 \pm 0.015$}}%
\newcommand{\Jtess}{\mbox{$0.792 \pm 0.012$}}%
\newcommand{\Jspec}{\mbox{$0.813 \pm 0.011$}}%
\newcommand{\Jsaao}{\mbox{$0.629 \pm 0.020$}}%
\newcommand{\thirdngts}{\mbox{$0.154\pm 0.003$}}%
\newcommand{\thirdtess}{\mbox{$0.249 \pm 0.003$}}%
\newcommand{\thirdspec}{\mbox{$0.167  \pm 0.003$}}%
\newcommand{\thirdsaao}{\mbox{$0.131 \pm 0.002$}}%
\newcommand{\semimaj}{\mbox{$3.8239\pm 0.0099$}}%
\newcommand{\orbinc}{\mbox{$84.76\,^{+0.19}_{-0.16}$}}%
\newcommand{\eccen}{\mbox{$0.00052\,^{+0.00130}_{-0.00034}$}}%
\newcommand{\longperi}{\mbox{$272\,^{+32}_{-215}$}}%
\newcommand{\prisyncvel}{\mbox{$18.60 \pm 0.22$}}%
\newcommand{\secsyncvel}{\mbox{$12.71 \pm 0.25$}}%

\defcitealias{GAIA2018}{G18}
\defcitealias{Zhang2020}{Z20}